\documentclass[12pt]{article}
\begin{document}
\begin{titlepage}
\begin{flushright}
hep-th/0209004 \\
TIT-HEP-483\\
August, 2002
\end{flushright}
\vspace{0.5cm}
\begin{center}
{\Large \bf 
Penrose Limit and String Theories \\on Various Brane Backgrounds
}
\lineskip .75em
\vskip2.5cm
{\large Hiroyuki Fuji},\ \  {\large Katsushi Ito}\ \  and \ \
{\large Yasuhiro Sekino}
\vskip 1.5em
{\large\it Department of Physics\\
Tokyo Institute of Technology\\
Tokyo, 152-8551, Japan}  
\vskip 3.5em
\end{center}
\begin{abstract}
We investigate the  Penrose limit of various brane solutions including 
D$p$-branes, NS5-branes, fundamental strings, 
$(p,q)$ fivebranes
and $(p,q)$ strings.
We obtain special null geodesics with the fixed radial coordinate (critical
radius),
along which the Penrose limit gives string theories with constant mass.
We also study string theories  with
time-dependent mass, which arise from the Penrose limit of the 
brane backgrounds.
We examine equations of motion of the strings 
in the asymptotic flat region and around the critical radius.
In particular, for $(p,q)$ fivebranes, 
we find that the string equations of motion in the
directions with the $B$ field are explicitly solved by the spheroidal
wave functions.
\end{abstract}
\end{titlepage}
\baselineskip=0.7cm
\catcode`\@=11
\@addtoreset{equation}{section}
\def\theequation{\thesection.\arabic{equation}}
\catcode`@=12
\relax
\section{Introduction}
The Penrose limits \cite{Pe,Gu} of backgrounds of the M theory and type II
superstring 
theories are useful 
for studying the holography between string theories and conformal
field theories \cite{BlFiPa,BeMaNa}.
This is based on the fact that the type IIB
Green-Schwarz superstring on the pp-wave background is solved exactly
in the light-cone gauge \cite{Me,MeTs}.

Superstring theories on various pp-wave 
backgrounds have 
been studied.
It would be quite interesting to study nonconformal or nonlocal
field theories in this framework 
\cite{GuNuMaSc,ZaSo,HuRaVe,AlKu2,BiKuPa,AlPa,HaSe}. 
Recently, the Penrose limit of 
the Pilch-Warner solution \cite{CoHaKeWa,GiPaSo,BrJoLoMy} 
and the D$p$-brane background in the near-horizon
limit \cite{GiPaSo} has been investigated from the viewpoint of the 
holographic RG flow.
In order to study string theories in such nontrivial backgrounds, one
must solve string theory in a time-dependent background.
Previous attempts to study
the string theories on the solvable plane wave
background can be seen in refs. \cite{AmKl,HoSte,VeSa,JoNu,KiKoLu}.
Recently, a solvable model based on $N=2$ supersymmetric sine-Gordon model 
was  proposed by Maldacena and Maoz \cite{MaMa}.

The purpose of this paper is to investigate the solvable string theories
in the pp-wave background with time-dependent mass.
We study the Penrose limit of various brane backgrounds including 
D$p$ ($p\leq 6$) branes, the NS5-branes and the fundamental strings.
We also
investigate the Penrose limit of the $(p,q)$ fivebranes and the 
$(p,q)$ strings, which are obtained by 
applying the $SL(2,{\bf Z})$ symmetry of type IIB superstrings.

In the present work, we do not restrict our analysis to  
the near-horizon limit.
The string theories on the Penrose limit of brane backgrounds include 
some interesting features.
When we consider the Penrose limit along the generic null geodesic, 
we have the flat space at infinity.
String theories on this background have time-dependent masses, 
which vanish in the asymptotic region.
We are also able to find special null geodesic for each brane solution 
such that the radial transverse 
coordinate from the brane is fixed. 
The Penrose limit along this geodesic with such a critical radius
gives rise to the Cahen-Wallach
space,
on which string theories have constant masses in the light-cone gauge.
This type of solution has been recently studied by Oz and Sakai 
\cite{OzSa} in the case of near-horizon limit of $(p,q)$ fivebranes.

We study the equations of motion of bosonic
strings in the plane wave background
with time-dependent masses, which reduce to the second-order ordinary
differential equations with respect to the radial transverse coordinate. 
We examine these differential equations 
for various brane solutions.
In the near horizon limit, a large class of backgrounds are shown to
be solved by the Bessel functions.
When we do not take the near horizon limit, the equations are difficult
to solve in general. 
In this paper, we study the case of 
$(p,q)$ fivebrane backgrounds in detail, since  we can solve 
the equations of motion in two regions: near the critical
radius and far from the origin.
The solutions are given by the Mathieu functions for both regions.
Furthermore, we are able to find the non-trivial exact solution 
in the directions with the $B$-field by using the spheroidal wave functions.

This paper is organized as follows:
In section 2, the Penrose limit of various $p$-brane solutions will be
studied.
The metric and  the $(p+2)$-form field strength 
are calculated as functions of the radial 
transverse coordinate.
In section 3, we study special null geodesics 
with the critical radius,
along which the Penrose limit gives the Cahen-Wallach space.
In section 4, we study the closed bosonic string theories on the
Penrose limit of various brane backgrounds.
In particular, we will consider the theories at the critical radius 
and in the near horizon limit.
In section 5, we will study the string theories on the $(p,q)$
fivebrane background. 
We examine equations of motion of the strings 
in the asymptotic flat region and around the critical radius.
We solve the string equation of motion in the directions with the $B$-field
using the spheroidal wave function.
In the Appendix, the string theory which is solvable in terms of the Bessel 
functions is discussed in some detail.

\section{Penrose limits of brane solutions}   

The Penrose limit is defined as a specific scaling of the metric and 
supergravity fields along a null geodesic.
In this section, we will review the Penrose limits of 
the $p$-brane solutions and give explicit forms of various brane solutions.

\subsection{Review of the Penrose limit}

We will briefly review the Penrose limit of the 
$p$-brane solution \cite{BlFiPa}.
The typical $p$-brane solution for a $D$-dimensional supergravity is 
\begin{eqnarray}
&&ds^2=A^2(r)ds^2({\bf E}^{1,p})+B^2(r)ds^2({\bf E}^{D-p-1}), \nonumber \\
&&F_{p+2}=dC_{p+1}=d{\rm  vol}({\bf E}^{1,p})\wedge dC(r), \nonumber \\
&&\phi=\phi(r)
\label{p-brane solution}
\end{eqnarray}
where $r$ is the radial transverse coordinate, ${\bf E}^{1,p}$ is the 
world-volume of the $p$-brane and ${\bf E}^{D-p-1}$ is the transverse space. 
$C_{p+1}$ is the $(p+1)$-form potential coupled to the brane.
$A$, $B$, $C$ and scalar field $\phi$ depend on $r$.
The $p$-brane metric is rewritten as follows.
\begin{eqnarray}
ds^2&=&A^2(r)(-dt^2+ds^2({\bf E}^p))
+B^2(r)\left(dr^2+r^2(d\psi^2+(\sin\psi)^2d\Omega^2_{D-p-3})\right)
\label{p-brane metric}
\end{eqnarray}
where $\psi$ is colatitude of $D-p-2$-dimensional sphere 
and $d\Omega^2_{D-p-3}$ is the metric for $D-p-3$ dimensional sphere.

We will consider the null geodesic in  $(t,r,\psi)$ space. 
With the metric 
\begin{eqnarray}
ds^2_{(3)}=-A^2dt^2+B^2dr^2+B^2r^2d\psi^2.
\label{trp metric}
\end{eqnarray}
There are two Killing vectors  
$\partial_t$ and $\partial_{\psi}$.
The corresponding conserved quantities are 
\begin{eqnarray}
E\equiv -g_{tt}{\dot t}=A^2{\dot t}, \quad 
J\equiv g_{\psi\psi}{\dot \psi}=B^2r^2{\dot \psi}
\label{conservation 1}
\end{eqnarray}
where the dot is the derivative with respect to the
affine parameter $\lambda$ along the geodesic.
$E$ and $J$ are energy and angular momentum respectively.
To normalize the energy $E=1$, 
we define an affine parameter $u\equiv E\lambda$.
The equations (\ref{conservation 1}) become
\begin{eqnarray}
A^2\frac{\partial t}{\partial u}=1,\quad
B^2r^2\frac{\partial \psi}{\partial u}= J/E.
\label{conservation}
\end{eqnarray}
The remaining free parameter which determines the geodesic is
$\ell\equiv J/E$.
The condition for a null geodesic is
\begin{eqnarray}
g_{tt}\left(\frac{\partial t}{\partial u}\right)^2
+g_{rr}\left(\frac{\partial r}{\partial u}\right)^2
+g_{\psi\psi}\left(\frac{\partial \psi}{\partial u}\right)^2
=B^2\left({\partial r\over \partial u}\right)^2
-\left(\frac{1}{A^2}-\frac{\ell^2}{B^2r^2}\right)
=0.
\label{EOM}
\end{eqnarray}
For a fixed $\ell$,  $r$ must satisfy $\ell\le rB/A$.
We will examine this inequality for each brane solution in section $3$.
When $\ell=0$, the null geodesic exists in $(t,r)$ space.
This is called radial null geodesic.

We introduce the coordinate transformation $(t,r,\psi)\to (u,v,\tilde{z})$
\begin{equation}
u=u(r), \quad v=t+\ell\psi+a(r), \quad \tilde{z}=\psi +b(r),
\label{eq:uvz}
\end{equation}
such that the metric (\ref{trp metric}) becomes
\begin{equation}
ds_{(3)}^2=2dudv+Kdv^2+Ldvd\tilde{z}+Md\tilde{z}^2. 
\label{metric (u,v,z)}
\end{equation}
In this metric, the tangent vector $\partial_u$ is a null geodesic vector.
The conditions for the null geodesic (\ref{conservation}) and 
(\ref{EOM}) lead to
\begin{eqnarray}
&&\frac{da}{dr}=\pm\sqrt{\frac{B^2}{A^2}-\frac{\ell^2}{r^2}}, \quad
\frac{db}{dr}=\mp\frac{\ell/r^2}{\sqrt{\frac{B^2}{A^2}-\frac{\ell^2}{r^2}}},
\nonumber \\
&&\frac{du}{dr}=\pm \frac{B^2}{\sqrt{\frac{B^2}{A^2}-\frac{\ell^2}{r^2}}}   
\equiv Q(r).  
\label{Q(r)} 
\end{eqnarray}
Then the functions $K$, $L$ and $M$ in the metric (\ref{metric (u,v,z)}) 
are given by
\begin{eqnarray}
K=-A^2, \quad L=2\ell A^2, \quad M=B^2r^2-\ell^2A^2.
\end{eqnarray} 
We rescale the coordinates,
the metric $g$, the  $(p+2)$-form field strength $F_{p+2}$ 
and the scalar field $\phi$.
\begin{eqnarray}
&&u\to u,\quad v\to \Omega^2 v, \quad \tilde{z}\to \Omega \tilde{z}, 
\quad \tilde{x}^a\to \Omega \tilde{x}^a, 
\quad \tilde{y}^{l}\to \Omega \tilde{y}^{l}
\nonumber \\
&&
g\to \Omega^{-2} g, \quad 
F_{p+2}\to \Omega^{-p}F_{p+2}, \quad \phi\to \phi
\end{eqnarray} 
where $\tilde{x}^a$ $(a=1,\cdots ,p)$ are the spatial coordinates of
the $p$-branes and $\tilde{y}^l$ $(l=1,\cdots ,D-p-3)$ 
are coordinates of $S^{D-p-3}$.
The Penrose limit is the limit where 
$\Omega\to 0$.
Then the $D$-dimensional metric 
and the $(p+2)$-form field strength $F_{p+2}$ become
\begin{eqnarray}
&&ds^2=2dudv+(B^2r^2-\ell^2A^2)dz^2+A^2ds^2({\bf E}^p)
+B^2r^2(\sin b)^2 ds^2({\bf E}^{D-p-3}), \nonumber \\
&&F_{p+2}=\pm \frac{dC(r)}{dr}\frac{\ell}{B}
\sqrt{\frac{1}{A^2}-\frac{\ell^2}{B^2r^2}}
du\wedge d\tilde{x}^1 \wedge \cdots \wedge d\tilde{x}^p\wedge d\tilde{z}.
\label{Rosen}
\end{eqnarray}
This coordinate is called the Rosen coordinate.

We transform the Rosen coordinate $(u,v,\tilde{x}^a,\tilde{y}^l)$ to
the Brinkman coordinate $(x^{\pm}, x^a,y^l)$
\begin{eqnarray}
&&u=x^{+}, \nonumber \\
&&v=x^{-}+\frac{\partial_+A(x^{+})}{2A(x^{+})}\sum_a(x^a)^2
+\frac{\partial_+\bigl(r(x^{+})B(x^{+})\sin b(x^+)\bigr)}
{2r(x^{+})B(x^{+})\sin b(x^+)}
\sum_l(y^l)^2
\nonumber \\
&&\quad\quad
+\frac{\partial_+\sqrt{B(x^{+})^2r(x^{+})^2-\ell^2A(x^{+})^2}}
{2\sqrt{B(x^{+})^2r(x^{+})^2-\ell^2A(x^{+})^2}}z^2,
\nonumber \\
&&\tilde{x}^a=\frac{x^a}{A(x^{+})},\nonumber \\
&&\tilde{y}^l=\frac{y^l}{r(x^{+})B(x^{+})}, \nonumber \\
&&\tilde{z}=\frac{z}{\sqrt{B^2(x^{+})r^2(x^{+})-\ell^2A^2(x^{+})}}.
\end{eqnarray}
In the Brinkman coordinate, the solution (\ref{Rosen}) becomes 
\begin{eqnarray}
&&ds^2=2dx^{+}dx^{-}+\left(
m_x^2\sum_{a=1}^{p}x_a^2 +m_y^2\sum_{l=1}^{D-p-3}y_l^2
+m_z^2z^2
\right)(dx^+)^2
\nonumber \\
&&\quad\quad\quad
+ds^2({\bf E}^{p})+ds^2({\bf E}^{D-p-3})+dz^2,
\nonumber \\
&&m_x^2(x^+)=\frac{\partial_+^2 A}{A},
\quad
m_y^2(x^+)=\frac{\partial_+^2(rB\sin b)}{rB\sin b},
\nonumber \\
&&m_z^2(x^+)=\frac{\partial_+^2\sqrt{B^2r^2-\ell^2A^2}}{\sqrt{B^2r^2-\ell^2A^2}},
\nonumber \\
&&F_{p+2}=\pm \frac{dC(r)}{dr}\frac{\ell}{rA^{p+1}B^2}
dx^{+}\wedge dx^{1}\wedge \cdots \wedge dx^{p}\wedge dz
\label{Brinkman}
\end{eqnarray}
where $ds^2({\bf E}^{p})=\sum_{a=1}^{p}(dx^a)^2$ and  
$ds^2({\bf E}^{D-p-3})=\sum_{l=1}^{D-p-3}(dy^l)^2$.

From (\ref{Q(r)}) and (\ref{Brinkman}), 
$m_i$ are written as functions of the coordinate $r$;
\begin{eqnarray}
&&m_x^2=\frac{-Q^{-3}\partial_rQ\partial_rA+Q^{-2}\partial_r^2A}{A},
\nonumber \\
&&m_y^2=\frac{-Q^{-3}\partial_rQ\partial_r(Br\sin b)
+Q^{-2}\partial_r^2(Br\sin b)}{Br\sin b},
\nonumber \\
&&m_z^2=\frac{-Q^{-3}\partial_rQ\partial_r\sqrt{B^2r^2-\ell^2A^2}
+Q^{-2}\partial_r^2\sqrt{B^2r^2-\ell^2A^2}}{\sqrt{B^2r^2-\ell^2A^2}}.
\label{WS mass}
\end{eqnarray}

\subsection{Plane wave geometry for various brane solutions}

We have obtained the plane wave  metric 
for typical $p$-brane solution.
In this subsection, we will write down the explicit forms 
of the plane wave
metric for the D$p$-brane ($p\le 6$), fundamental string,
NS$5$-brane, $(p,q)$ string and $(p,q)$ fivebrane solutions. 
We also consider the near horizon limit.
Since we are interested in the string theory on these backgrounds,  
the metrics which are discussed below are those in the string frame.

\vspace{2mm}
\noindent{{\bf ($1$) D$p$-brane solution}}
\vspace{1mm}

First we discuss the Penrose limit of the D$p$-branes.
In string frame, the D$p$-brane solution takes the form
\cite{HoSt}
\begin{eqnarray}
&&ds^2=H^{-1/2}ds^2({\bf E}^{1,p})+H^{1/2}ds^2({\bf E}^{9-p}), \nonumber  \\
&&F_{p+2}=d{\rm vol}({\bf E}^{1,p})\wedge dH^{-1}, \nonumber \\
&&e^{2\phi}=H^{\frac{3-p}{2}}.
\end{eqnarray}
$H(r)$ is the harmonic function on ${\bf E}^{9-p}$
\begin{eqnarray}
H=1+\frac{Q_p}{r^{7-p}}, \quad Q_{p}=c_{p}g_{YM}^{2}N(\alpha')^{5-p}
\end{eqnarray}
where $c_{p}=2^{7-2p}\pi^{{9-3p\over2}}\Gamma({7-p\over2})$ and
$g_{YM}^2=(2\pi)^{p-2}g_{s}(\alpha')^{{p-3\over2}}$. 
See e.g. ref. \cite{IzMaSoYa}.

In this case, $A(r)=H^{-1/4}$, $B(r)=H^{1/4}$ and $C(r)=H^{-1}$.
Plugging these into (\ref{Q(r)}),
we obtain\footnote{
In the following we will choose 
the upper signs in (\ref{Q(r)}).
}
\begin{eqnarray}
&&\frac{db}{dr}=-\frac{\ell}{r\sqrt{r^2-\ell^2+Q_pr^{-5+p}}}, \nonumber
\\
&&Q(r)=\frac{du}{dr}=\left({r^{7-p}+Q_{p}\over r^{7-p}+Q_{p}
-\ell^2 r^{5-p}}\right)^{1\over2}.
\label{QDp}
\end{eqnarray}
From (\ref{Brinkman}), 
we find the Penrose limit of the metric and the $(p+2)$-form field strength
\begin{eqnarray}
&&ds^2=2dx^+dx^-+\left(
m_x^2\sum_{a=1}^{p}x_a^2 +m_y^2\sum_{l=1}^{7-p}y_l^2
+m_z^2dz^2
\right)(dx^+)^2
+ds^2({\bf E}^{8}),\nonumber \\
&&
F_{p+2}=\ell (7-p)Q_pr^{9-p}(1+Q_pr^{7-p})^{\frac{p-9}{4}}
dx^+\wedge dx^1\wedge \cdots \wedge dx^p\wedge dz.
\end{eqnarray}
The explicit forms of $m_i$ are given by (\ref{WS mass})
\begin{eqnarray}
&&m_x^2=(7-p)Q_{p}\Biggl[
(3-p)Q_{p}^2+(3p-13)Q_{p}\ell^2 r^{5-p}-(29-3p)Q_{p}r^{7-p}
\nonumber \\
&&\quad\quad\quad
+4(9-p)\ell^2 r^{12-2p}-4(8-p)r^{14-2p}
\Biggr]\bigg/ 16 r^2(r^{7-p}+Q_{p})^3,
\nonumber \\
&&m_y^2=(7-p)Q_{p}\Biggl[
(3-p)Q_{p}^2-(p+1)Q_{p}\ell^2 r^{5-p}+(27-5p)Q_{p}r^{7-p}
\nonumber \\
&&\quad\quad\quad
-4(9-p)\ell^2 r^{12-2p}+4(6-p)r^{14-2p}
\Biggr]\bigg/ 16 r^2(r^{7-p}+Q_{p})^3,
\nonumber \\
&&m_z^2=(7-p)Q_{p}\Biggl[
(3-p)Q_{p}^2+(3p-13)Q_{p}\ell^2 r^{5-p}+(27-5p)Q_{p}r^{7-p}
\nonumber \\
&&\quad\quad\quad
+4(9-p)\ell^2 r^{12-2p}+4(6-p)r^{14-2p}
\Biggr]\bigg/ 16 r^2(r^{7-p}+Q_{p})^3.
\label{eq:dpmass}
\end{eqnarray}
In the near horizon limit, these $m_i$ become
\begin{eqnarray}
&&m_{z}^{2}
=m_{x}^2=
{(7-p)\left[
(3-p)Q_{p}+(3p-13)\ell^2 r^{5-p}
\right]\over 16 r^2 Q_{p}},
\nonumber\\
&&m_{y}^{2}
=
{(7-p)\left[
(3-p)Q_{p}+(-1-p)\ell^2 r^{5-p}
\right]\over 16 r^2 Q_{p}},
\label{eq:nhDpmass}
\end{eqnarray}
and the $(p+2)$-form field strength becomes 
\begin{eqnarray}
&&
F_{p+2}=\ell (7-p)Q_p^{\frac{p-5}{4}}r^{(p-3)(p-9)/4}dx^{+}\wedge
dx^1\wedge \cdots \wedge dx^{p}\wedge dz,
\end{eqnarray}
which agree with the result in ref. \cite{GiPaSo}.

\vspace{2mm}
\noindent{{\bf ($2$) NS$5$-brane solution}}
\vspace{1mm}

Next we will discuss the Penrose limit of the NS$5$-branes.
The NS$5$-brane solution is
\cite{CaHaSt}
\begin{eqnarray}
&&ds^2=ds^2({\bf E}^{1,5})+H ds^2({\bf E}^{4}), \nonumber \\
&&F_{7}=d{\rm vol}({\bf E}^{1,5})\wedge dH^{-1}, \quad
e^{2\phi}=H,  \nonumber \\
&&H=1+\frac{Q_5}{r^2}. 
\end{eqnarray}
Since we are working in string frame,
the dual NS-NS $3$-form field strength $H_3=dB_2$ is
\begin{eqnarray}
&&H_3=e^{2\phi}*F_7=2Q_5d{\rm vol}(S^3)
\end{eqnarray}
where $S^3$ is in the transverse space ${\bf E}^4$.

In this case, since $A(r)=1$, $B(r)=H^{1/2}$ and $C(r)=H^{-1}$,
$du/dr$ becomes
\begin{eqnarray}
\frac{du}{dr}=Q(r)=\frac{r^2+Q_{5}}{r\sqrt{r^2+Q_{5}-\ell^2}}.
\label{QNS5}
\end{eqnarray}
The Penrose limit of the metric and the NS-NS $3$-form field strength $H_3$ are
\begin{eqnarray}
&&ds^2=2dx^+dx^-+\left(
m_x^2\sum_{a=1}^{5}x_a^2 +m_y^2\sum_{l=1}^{2}y_{l}^2
+m_z^2dz^2
\right)(dx^{+})^2
+ds^2({\bf E}^{8}),\nonumber \\
&&
H_{3}= \frac{2\ell Q_5}{(r^2+Q_5)^2}
dx^+\wedge dy^1\wedge dy^2
\end{eqnarray}
where
\begin{eqnarray}
m^{2}_{x}&=&0, \nonumber\\
m^2_{y}&=& {Q_{5}(-Q_{5}\ell^2+2r^2 Q_{5}-4r^2\ell^2+2r^4)\over 
(r^2+Q_{5})^4},\nonumber\\
m^{2}_{z}&=& {2 Q_{5}r^2\over (r^2+Q_{5})^3}.
\label{eq:NS5mass}
\end{eqnarray}
In the near horizon limit, $m_i$ become constants
\begin{eqnarray}
&&m^{2}_{x}=0,\quad 
m^2_{y}= -{\ell^2\over Q_{5}^2},\quad
m^{2}_{z}= 0,\label{NHL:NS5mass}
\end{eqnarray}
and the NS-NS $3$-form field strength is
\begin{eqnarray}
&&H_{3}= \frac{2\ell}{{Q_5}}dx^+\wedge dy^1\wedge dy^2.
\end{eqnarray}
For $\ell \ne 0$,
this plane wave geometry becomes the Nappi-Witten 
geometry \cite{NaWi}
as noted in refs. \cite{Go-Oo,KiPi}.
Taking the Penrose limit along the radial null geodesic $\ell=0$, 
all of $m_i$ become zero and 
the resulting string theory is linear dilaton theory \cite{BlFiPa}.

\vspace{2mm}
\noindent{{\bf ($3$) Fundamental string solution}}
\vspace{1mm}

The fundamental string solution is
\cite{DaGiHaRu}
\begin{eqnarray}
&&ds^2=H^{-1}ds^2({\bf E}^{1,1})+ ds^2({\bf E}^{8}), \nonumber \\
&&F_{3}=d{\rm vol}({\bf E}^{1,1})\wedge dH^{-1}, \quad
e^{2\phi}=H^{-1},
\nonumber \\
&&H=1+\frac{Q_1}{r^6}.
\end{eqnarray}

Since $A=H^{-1/2}$,  $B=1$ and $C=H^{-1}$ in this case, we obtain
\begin{eqnarray}
{du\over dr}=
Q(r)={r^3\over\sqrt{r^6+Q_{1}-\ell^2 r^4}}.
\label{QF1} 
\end{eqnarray}
The Penrose limit of the metric and the NS-NS $3$-form field strength $F_3$
are
\begin{eqnarray}
&&ds^2=2dx^+dx^-+\left(
m_x^2x^2 +m_y^2\sum_{l=1}^{6}y_{l}^2
+m_z^2dz^2
\right)(dx^+)^2
+ds^2({\bf E}^{8}),\nonumber \\
&&
F_{3}= \frac{6\ell Q_1}{r^2(r^6+Q_1)}
dx^+\wedge dx\wedge dz
\end{eqnarray}
where $m_i$ are
\begin{eqnarray}
m^{2}_{x}&=& -3 {Q_{1}(Q_{1}^2+Q_{1}\ell^2 r^4+8Q_{1}r^6-8\ell^2 r^{10}+
7r^{12})
\over r^8 (r^6+Q_{1})^2},  \nonumber\\
m^2_{y}&=& -3{Q_{1}\over r^8},\nonumber\\
m^{2}_{z}&=&-3 {Q_{1}(Q_{1}^2+Q_{1}\ell^2 r^4+2Q_{1}r^6-8\ell^2 r^{10}+r^{12})
\over r^8 (r^6+Q_{1})^2}.
\label{eq:F1mass}
\end{eqnarray}
In the  near horizon limit, $m_i$ become
\begin{eqnarray}
&&m^{2}_{x}= -3 {Q_{1}+\ell^2 r^4 \over r^8 },\quad
m^{2}_{y}= -3 {Q_{1}\over r^8},\quad
m^{2}_{z}= -3 {Q_{1}+\ell^2 r^4 \over r^8 },
\end{eqnarray}
and the NS-NS $3$-form field strength is
\begin{eqnarray}
&& F_3=\frac{6\ell}{r^2}dx^+\wedge dx\wedge dz.
\end{eqnarray}

\vspace{2mm}
\noindent{{\bf ($4$) $(p,q)$ fivebrane solution}}
\vspace{1mm}

Since type IIB superstring theory has S-duality, 
there is a $(p,q)$ fivebrane solution.
The metric and the $B$-field for the $(p,q)$ fivebranes are derived 
from those of the NS$5$-brane solution
by acting $SL(2,{\bf Z})$ duality transformation 
in Einstein frame \cite{LuRo,AlOz}.
By pulling back to the string frame, the metric, NS-NS $3$-form field 
strength $H_3$, 
RR $3$-form field strength $F_3$, 
dilaton $\phi$ and axion $\chi$
become 
\begin{eqnarray}
&&ds^2
=h^{-1/2}
(-dt^2+\sum_{a=1}^5dx_a^2 + f (dr^2+ r^2 d\Omega_{3}^2)), \nonumber \\
&&f=1+{\tilde{Q}_5\over  r^2},\quad
h^{-1}=\sin^2\gamma f^{-1} +\cos^2\gamma, 
\nonumber \\
&&
H_3=2\cos\gamma\tilde{Q}_5 d{\rm vol}(S^3),\quad
F_3=2\sin\gamma\tilde{Q}_5 d{\rm vol}(S^3),\nonumber \\
&&e^{2\phi}=fh^{-2}, \quad \chi=h\sin\gamma\cos\gamma (f^{-1}-1)
\label{eq:pq5solution}
\end{eqnarray}
where $\gamma$ is defined by
\begin{eqnarray}
\cos\gamma=\frac{p}{\sqrt{p^2+q^2}},\quad \sin\gamma=\frac{q}{\sqrt{p^2+q^2}}.
\label{def gamma}
\end{eqnarray}
When $\gamma=0$ and $\gamma=\pi/2$, 
this solution reduces to that of the NS$5$-branes and the D$5$-branes, 
respectively.

In this case, $A(r)=h^{-1/4}$ and $B(r)=h^{-1/4}f^{1/2}$.
The equation in (\ref{Q(r)}) leads to
\begin{eqnarray}
{du\over dr}=Q(r)
={\sqrt{r^2+\tilde{Q}_5}\sqrt{r^2+\cos^2\gamma \tilde{Q}_5}\over
r \sqrt{r^2+\tilde{Q}_5-\ell^2}}. 
\label{Q(p,q)5} 
\end{eqnarray}
Taking the Penrose limit, the resulting metric and 
the $3$-form field strengths are
\begin{eqnarray}
&&ds^2=2dx^+dx^-+\left(
m_x^2\sum_{a=1}^{5}x_a^2 +m_y^2\sum_{l=1}^{2}y_{l}^2
+m_z^2dz^2
\right)(dx^+)^2
+ds^2({\bf E}^{8}),\nonumber \\
&&
H_{3}= \frac{2\ell(\cos\gamma) \tilde{Q}_5 }
{(r^2+\tilde{Q}_5)(r^2+\tilde{Q}_5\cos^2\gamma)}
dx^+\wedge dy^1\wedge dy^2, \nonumber \\
&&
F_{3}= \frac{2\ell(\sin\gamma) \tilde{Q}_5}
{(r^2+\tilde{Q}_5)(r^2+\tilde{Q}_5\cos^2\gamma)}
dx^+\wedge dy^1\wedge dy^2
\end{eqnarray}
where $m_i$ become
\begin{eqnarray}
&&m^2_x(r)= {\sin^2\gamma \tilde{Q}_{5} r^2\over 4 (r^2+\tilde{Q}_{5})^3
(r^2+{\,\cos^2\!\gamma\,}\tilde{Q}_{5})^3}\Big[ 4{\,\cos^2\!\gamma\,}\tilde{Q}_{5}^2 (\tilde{Q}_{5} -\ell^2)  
\nonumber \\
&&\quad\quad\quad\quad
+\{(-1+ 3{\,\cos^2\!\gamma\,} )\tilde{Q}_{5}^2+(1+3{\,\cos^2\!\gamma\,})\tilde{Q}_{5}\ell^2 \}r^2
\nonumber \\
&&\quad\quad\quad\quad
+\{-(7+{\,\cos^2\!\gamma\,})\tilde{Q}_{5}+8\ell^2 \}r^4 -6 r^6\Big],
\label{pq5mx} \\
&&m^2_y(r)={\tilde{Q}_{5}\over 4 (r^2+\tilde{Q}_{5})^3
(r^2+{\,\cos^2\!\gamma\,}\tilde{Q}_{5})^3}\Big[ -4\cos^4\gamma\tilde{Q}_{5}^3 \ell^2\nonumber\\
&&
\quad\quad\quad\quad+ \{ 4{\,\cos^2\!\gamma\,} (1+{\,\cos^2\!\gamma\,})\tilde{Q}_{5}^3 -12{\,\cos^2\!\gamma\,}(1+{\,\cos^2\!\gamma\,})
 \tilde{Q}_{5}^2\ell^2\}r^2
\nonumber\\
&&\quad\quad\quad\quad
+ \{(-1+20{\,\cos^2\!\gamma\,} +5\cos^4\gamma)\tilde{Q}_{5}^2  -3(1+10{\,\cos^2\!\gamma\,} +\cos^4\gamma)\tilde{Q}_{5}\ell^2
\}r^4\nonumber\\
&&\quad\quad\quad\quad
+ \{ (1+22{\,\cos^2\!\gamma\,} +\cos^4\gamma)\tilde{Q}_{5}  -8(1+{\,\cos^2\!\gamma\,})\ell^2
\}r^6
+ 2(1+3{\,\cos^2\!\gamma\,})r^8\Big],
\nonumber \\
\label{pq5my}\\
&&m^2_z(r)={\tilde{Q}_{5} r^2\over 4 (r^2+\tilde{Q}_{5})^3
(r^2+{\,\cos^2\!\gamma\,}\tilde{Q}_{5})^3}
\Big[ 
 4{\,\cos^2\!\gamma\,} (1+{\,\cos^2\!\gamma\,})\tilde{Q}_{5}^3 
-4 {\,\cos^2\!\gamma\,}\sin^{2}\gamma\tilde{Q}_{5}^2\ell^2 
\nonumber \\
&&\quad\quad\quad\quad
+ \{(-1+20{\,\cos^2\!\gamma\,} +5\cos^4\gamma)\tilde{Q}_{5}^2
  +\sin^{2}\gamma(1+3{\,\cos^2\!\gamma\,})\tilde{Q}_{5}\ell^2
 \}r^2\nonumber\\
&&\quad\quad\quad\quad
+ \{ (1+22{\,\cos^2\!\gamma\,} +\cos^4\gamma)\tilde{Q}_{5}
  +8\sin^{2}\gamma\ell^2
\}r^4
+ 2(1+3{\,\cos^2\!\gamma\,})r^6\Big].
\label{pq5mz}
\end{eqnarray}
It is easily checked that 
they reduce to $m_i$ of the NS$5$-branes and the D$5$-branes 
when $\cos\gamma=1$ and $\cos\gamma=0$, respectively.

The near horizon limit of the $(p,q)$ fivebrane solution
is given by replacing $f=1+\tilde{Q}_{5}/r^2$ 
with $\tilde{Q}_{5}/r^2$ in (\ref{eq:pq5solution}).
Taking the Penrose limit of this background, 
$m_i$ become
\begin{eqnarray}
 m_{x}^2&\!=\!& m_{z}^2=
{1\over4} {r^2\sin^2\gamma (\tilde{Q}_{5}-\ell^2)
(-r^2\sin^2\gamma +4\tilde{Q}_{5}\cos^2\gamma)\over
\tilde{Q}_{5}(r^2\sin^2\gamma +\tilde{Q}_{5}\cos^2\gamma)^3 },
\nonumber  \\
m_{y}^2&=&
-{1\over4} {
4\tilde{Q}_{5}^2\ell^2\cos^4\gamma
-4\tilde{Q}_{5}\sin^2\gamma\cos^2\gamma(\tilde{Q}_{5}-3\ell^2)r^2
+\sin^4\gamma (\tilde{Q}_{5}+3\ell^2)r^4
\over
\tilde{Q}_{5}(r^2\sin^2\gamma +\tilde{Q}_{5}\cos^2\gamma)^3 },
\end{eqnarray}
and the 3-form field strengths become
\begin{eqnarray}
&&H_3=\frac{2\ell\cos\gamma}{\tilde{Q}_5\cos^2\gamma +r^2\sin^2\gamma}
dx^{+}\wedge dy^1\wedge dy^2,
\nonumber \\
&&F_3=\frac{2\ell\sin\gamma}{\tilde{Q}_5\cos^2\gamma +r^2\sin^2\gamma}
dx^{+}\wedge dy^1\wedge dy^2.
\end{eqnarray}

\vspace{2mm}
\noindent{{\bf ($5$) $(p,q)$ string solution}}
\vspace{1mm}

Finally we will consider the $(p,q)$ strings.
By $SL(2,{\bf Z})$ symmetry of type IIB superstrings, 
the metric and the $B$-field for the $(p,q)$ strings are also derived 
from those of the fundamental string solution in Einstein frame \cite{Sc}.
In string frame, 
the metric, NS-NS $3$-form field strength $F_3$, 
RR $3$-form field strength $H_3$,
dilaton $\phi$ and axion $\chi$
are
\begin{eqnarray}
&&ds^2=h^{-1/2}(
f^{-1}(-dt^2+dx^2) + dr^2+ r^2 d\Omega_{7}^2),\nonumber\\
&&f=1+{\tilde{Q}_1\over  r^6},\quad
h^{-1}=\sin^2\gamma f +\cos^2\gamma, 
\nonumber \\
&&
F_3=\cos\gamma d{\rm vol}({\bf E}^{1,1})\wedge df^{-1},\;
H_3=\sin\gamma d{\rm vol}({\bf E}^{1,1})\wedge df^{-1},
\nonumber
\\
&&e^{2\phi}=f^{-1}h^{-2},\quad \chi =h\sin\gamma\cos\gamma (f-1)
\label{(p,q)string}
\end{eqnarray}
where $\gamma$ is defined by (\ref{def gamma}) as the $(p,q)$ fivebranes.
The charge $\tilde{Q}_1$ is related with the original fundamental string 
charge $Q_1$ by $\tilde{Q}_1=\sqrt{p^2+q^2}Q_1$.
When $\gamma=0$ and $\gamma=\pi/2$, 
this solution reduces to the fundamental string and the D$1$-brane
solution, respectively.

Since $A(r)=h^{-1/4}f^{-1/2}$ and  $B(r)=h^{-1/4}$, 
we obtain
\begin{eqnarray}
{du\over dr}=Q(r)
={\sqrt{r^6+\sin^{2}\gamma \tilde{Q}_{1}}\over
\sqrt{r^6-\ell^2r^4+\tilde{Q}_{1}}}.
\label{Q(p,q)1} 
\end{eqnarray}
The Penrose limit of this solution is
\begin{eqnarray}
&&ds^2=2dx^+dx^-+\left(
m_x^2x^2 +m_y^2\sum_{l=1}^{6}y_{l}^2
+m_z^2dz^2
\right)(dx^+)^2
+ds^2({\bf E}^{8}),\nonumber \\
&&
F_{3}= \frac{6\cos\gamma\ell \tilde{Q}_1 r^4}
{(r^6+\tilde{Q}_1)(r^6+\tilde{Q}_1\sin\gamma)}
dx^+\wedge dx\wedge dz, \nonumber \\
&&
H_{3}= \frac{6\sin\gamma\ell \tilde{Q}_1 r^4}
{(r^6+\tilde{Q}_1)(r^6+\tilde{Q}_1\sin\gamma)}
dx^+\wedge dx\wedge dz
\end{eqnarray}
where we used $C=\cos\gamma f^{-1}$ and $C=\sin\gamma f^{-1}$ in 
$F_3$ and $H_3$, respectively.
In the above equations, $m_i$ are
\begin{eqnarray}
&&m^2_x(r)= -{3 \tilde{Q}_{1}\over 4 r^2(r^6+\tilde{Q}_{1})^2
(r^6+\sin^{2}\gamma\tilde{Q}_{1})^3}
\Big[ -\sin^{4}\gamma\tilde{Q}_{1}^4 
+5 \sin^{4}\gamma \tilde{Q}_{1}^3\ell^2 r^4
\nonumber \\
&&\quad\quad\quad\quad
-\sin^{2}\gamma(-11+23\cos^2\gamma )\tilde{Q}_{1}^3 r^6
+6\sin^{2}\gamma(-1+5\cos^2\gamma )\tilde{Q}_{1}^2\ell^2 r^{10}
\nonumber \\
&&\quad\quad\quad\quad
+(39-60\cos^2\gamma+25\cos^4\gamma)\tilde{Q}_{1}^2 r^{12}
+(-27+30\cos^2\gamma+\cos^4\gamma)\tilde{Q}_{1}\ell^2 r^{16}
\nonumber \\
&&\quad\quad\quad\quad
+(41-10\cos^2\gamma+\cos^4\gamma)\tilde{Q}_{1} r^{18}
-16(1+\cos^2\gamma)\ell^2 r^{22}
+14(1+\cos^2\gamma)r^{24}\Big],
\nonumber \\
&&m^2_y(r)= -{3 \tilde{Q}_{1}\over 4 r^2
(r^6+\sin^{2}\gamma\tilde{Q}_{1})^3}
\Big[ -\sin^{4}\gamma\tilde{Q}_{1}^2 + \sin^{4}\gamma \tilde{Q}_{1}\ell^2 r^4
-\sin^{2}\gamma(11+\cos^2\gamma )\tilde{Q}_{1} r^6
\nonumber \\
&&\quad\quad\quad\quad
+16\sin^{2}\gamma\ell^2 r^{10}
+(-10+14\cos^2\gamma)r^{12}\Big],
\nonumber\\
&&m^2_z(r)= -{3 \tilde{Q}_{1}\over 4 r^2(r^6+\tilde{Q}_{1})^2
(r^6+\sin^{2}\gamma\tilde{Q}_{1})^3}
\Big[ -\sin^{4}\gamma\tilde{Q}_{1}^4 
+5 \sin^{4}\gamma \tilde{Q}_{1}^3\ell^2 r^4
\nonumber \\
&&\quad\quad\quad\quad
+\sin^{2}\gamma(-13+\cos^2\gamma )\tilde{Q}_{1}^3 r^6
+6\sin^{2}\gamma(-1+5\cos^2\gamma )\tilde{Q}_{1}^2\ell^2 r^{10}
\nonumber \\
&&\quad\quad\quad\quad
+(-33+36\cos^2\gamma+\cos^4\gamma)\tilde{Q}_{1}^2 r^{12}
+(-27+30\cos^2\gamma+\cos^4\gamma)\tilde{Q}_{1}\ell^2 r^{16}
\nonumber \\
&&\quad\quad\quad\quad
+(-31+38\cos^2\gamma+\cos^4\gamma)\tilde{Q}_{1} r^{18}
-16(1+\cos^2\gamma)\ell^2 r^{22}
\nonumber \\
&&\quad\quad\quad\quad
+(-10+14\cos^2\gamma)r^{24}\Big].\label{pq1mz}
\end{eqnarray}
It is also checked that they become the Penrose limit of 
the fundamental string solution and the D$1$-brane solution 
when $\cos\gamma=1$ and $\cos\gamma=0$, respectively.

In the near horizon limit, which is obtained by
the replacement $f=1+\tilde{Q}_{2}/r^{6} \rightarrow
\tilde{Q}_{2}/r^6$ in (\ref{(p,q)string}), $m_i$ become
\begin{eqnarray}
&& m_{x}^2= m_{z}^2=
-{3\over4} {1
\over
r^2 (r^6\cos^2\gamma +\tilde{Q}_{1}\sin^2\gamma)^3 }
\biggl[4\ell^2r^{16}\cos^4\gamma
+4\tilde{Q}_{1} r^{12}\cos^4\gamma
\nonumber\\
&&\quad\quad\quad\quad\quad\quad
+24\tilde{Q}_{1}\ell^2r^{10}\cos^2\gamma\sin^2\gamma
-12\tilde{Q}_{1} r^6 \cos^2\gamma\sin^2\gamma
\nonumber \\
&&\quad\quad\quad\quad\quad\quad
+5\tilde{Q}_{1}^{2}\ell^2 r^4\sin^4\gamma 
-\tilde{Q}_{1}^{3}\sin^4\gamma
\biggr], \nonumber\\
&&m_{y}^2=
-{3\over4} {1
\over
r^2 (r^6\cos^2\gamma +\tilde{Q}_{1}\sin^2\gamma)^3 }
\tilde{Q}_{1}\biggl[
4r^{12}\cos^4\gamma+16\ell^2r^{10}\cos^2\gamma\sin^2\gamma
\nonumber\\
&&\quad\quad\quad
-12\tilde{Q}_{1}r^6 \cos^2\gamma\sin^2\gamma
+\tilde{Q}_1\ell^2r^4\sin^4\gamma-\tilde{Q}_{1}^2\sin^4\gamma
\biggr],
\end{eqnarray}
and  the $3$-form field strengths are
\begin{eqnarray}
&&F_3=
\frac{6\ell r^4 \cos\gamma}{r^6\cos^2\gamma +\tilde{Q}_1\sin^2\gamma}
dx^+\wedge dx\wedge dz,
\nonumber \\
&&H_3=
\frac{6\ell r^4 \sin\gamma}{r^6\cos^2\gamma +\tilde{Q}_1\sin^2\gamma}
dx^+\wedge dx\wedge dz.
\label{NHLpq1}
\end{eqnarray}

\section{Null geodesics and critical radius}


In the previous section, we have 
obtained the Penrose limits of
various brane solutions. 
In this section, we examine the allowed region
for the radial coordinate $r$ of a null geodesic 
with a given parameter $\ell$ for each brane solution. 
In particular, we point out the existence of 
special null geodesics which stay at a 
fixed (`critical') radius. The Penrose limits
along these geodesics are of interest, 
since, as we will see in section 4, 
the string theories on these backgrounds 
have constant masses in the light-cone gauge
and are solvable. We will present explicit forms
of the plane wave geometry resulting from such
Penrose limits.

We begin with the discussion on the geodesics
for the D$p$-brane solutions.
The relation between the affine parameter $u$ 
and the radial coordinate $r$
is given by (\ref{QDp}) for the D$p$-branes. 
Since $du/dr$ is necessarily real, 
the region of $r$ must obey
\begin{equation}
f(r)\equiv r^{7-p}+Q_{p}-\ell^{2}r^{5-p}\ge 0.
\label{fr}
\end{equation}

We shall first consider the condition that a null geodesic
starting from $r=\infty$ reach the origin $r=0$.
It is given by $f(r)> 0$ throughout $0\le r\le \infty$, 
that is,
\begin{equation}
f(r_{min})>0
\label{frmin}
\end{equation}
where $f(r_{min})$ is the minimum of $f(r)$ in the
region $0\le r\le \infty$.
When $p\le 4$, as we can see from (\ref{fr}),  
the minimum of $f(r)$ is at
\begin{equation}
r_{min}=\left({5-p\over 7-p}\right)^{1/2}\ell.
\end{equation}
Thus, (\ref{frmin}) amounts to
\begin{equation}
\ell< (c Q_{p})^{1\over 7-p} \quad \mbox{where}\quad 
c={7-p\over 2}\left({7-p\over 5-p}\right)^{5-p\over 2}
\qquad (p\le4).
\label{geodcond}
\end{equation}
In the cases of $p=5,6$ the minimum of $f(r)$ is at $r=0$.
For $p=5$, $f(0)=Q_{5}-\ell^2$ and the condition (\ref{frmin})
becomes
\begin{equation}
Q_{p}> \ell^2. 
\end{equation}
For $p=6$, the condition is 
\begin{equation}
\ell=0.
\end{equation}

If (\ref{frmin}) is not the case, 
the equation $f(r)=0$ have positive root(s). 
Let us first consider the $p\le 4$ cases. 
When (\ref{geodcond}) is not satisfied, there
are two positive roots $r_{0\pm}$ 
($r_{0+}\ge r_{0-}$). 
Explicit forms of $r_{0\pm}$ for $p=1,3,4$ are
found by solving second or third order equations
and given as follows. 
For $p=1$, and when $27Q_1/4\le \ell^6$, we have
\begin{eqnarray*}
&&\hspace{-0.5cm} r_{0\pm}=\Biggl[
\frac{\ell^2}{3}-\biggl\{
\left(1\pm i\sqrt{3}\right)
\left(2\ell^6-27Q_1- i3\sqrt{3}\sqrt{4\ell^6Q_1-27Q_1^2}\right)^{1/3}
\nonumber \\
&&\;\; 
+\left(1\mp i\sqrt{3}\right)\left(2\ell^6-27Q_1+ i3\sqrt{3}\sqrt{4\ell^6Q_1-27Q_1^2}\right)^{1/3}
\biggr\}
\Big/
(6\cdot 2^{1/3})
\Biggr]^{1/2}.
\end{eqnarray*}
For $p=3$, when $4Q_3\le \ell^4$, 
\begin{eqnarray*}
&&r_{0\pm}=\sqrt{\frac{\ell^2\pm\sqrt{\ell^4-4Q_3}}{2}}.
\end{eqnarray*}
For $p=4$, when $27Q_{4}^{2}/4\le \ell^{6}$,
\begin{eqnarray*}
&&r_{0\pm}=
\Biggl\{
\left(1\mp i\sqrt{3}\right)\left(-27Q_4+i3\sqrt{12\ell^{6}-81Q_4^2}\right)^{1/3}
\nonumber \\
&&\quad \quad\quad
+\left(1\pm i\sqrt{3}\right)\left(-27Q_4-i3\sqrt{12\ell^{6}-81Q_4^2}\right)^{1/3}
\Biggr\}\Bigg/(6\cdot 2^{1/3}).
\end{eqnarray*}
For the cases under consideration, a null geodesic  
which starts at $r=\infty$ can reach only to $r=r_{0+}$, 
but no nearer to the origin.
We may also consider a null geodesic which remain 
in the region near the origin $0\le r\le r_{0-}$
where $f(r)>0$.
In addition, there are special null geodesics which stay
at a certain fixed radius $r=r_{0+}$ or $r=r_{0-}$,
since $dr/du=0$  at these points\footnote{%
To see the behavior of the geodesic around
the radius where $dr/du=0$ more precisely, we need to
study the geodesic equations and examine the quantities 
such as the second derivative with respect to the affine 
parameter.}.

In the cases of $p=5,6$,
on the other hand, the equation $f(r)=0$ can have only
one positive root $r_{0}$. For $p=5$, when $\ell^2\le Q_5$,
the root is  
\begin{equation}
r_0=\sqrt{\ell^2-Q_5}.
\end{equation}
and for $p=6$, when $\ell\neq 0$,
\begin{eqnarray}
r_{0}=\frac{\sqrt{Q_6^2+4\ell^2}-Q_6}{2}.
\end{eqnarray}
In these cases,
a geodesic which starts at $r=\infty$ can reach
only to $r=r_0$. We can also consider a geodesic which
stays at $r=r_0$. However, there is no geodesic which
remain within finite radius near the origin, in contrast
to the $p\le 4$ case. 

The conditions for the allowed region of $r$ for
the NS and $(p,q)$ fivebranes are the same as 
the one for the D5-branes (with $Q_5$ replaced
by $\tilde{Q}_{5}$). It is because the reality conditions
of $du/dr$ for these backgrounds
are the same as the one for the D5-branes,
i.e. (\ref{fr}) with $p=5$, as we can see from 
(\ref{QNS5}) and (\ref{Q(p,q)5}).
For the same reason, the conditions for the allowed
region of $r$ for the fundamental and $(p,q)$ strings
are the same as the one for the D1-branes (with
$Q_1$ replaced by $\tilde{Q}_{1}$).

When we consider the near-horizon geometries of
the brane solutions, there are slight modifications
to the above discussion.
The derivative $du/dr$ becomes in the near-horizon
limit 
\[
 {du\over dr}=\left(Q_{p}\over Q_{p}-\ell^2 r^{5-p}\right)^{1\over2}.
\]
For $p\neq 5$, there is one solution of
$dr/du=0$ ($du/dr=\infty$):
\begin{equation}
r_0=\left({Q_{p}\over\ell^2}\right)^{1\over 5-p} \qquad (p\ne 5).
\label{eq:nhr0}
\end{equation}
The allowed region of the geodesic for $p\le 4$
is $0\le r \le r_0$. 
On the other hand, for $p=6$, the allowed region
is $r_0\le r$, and the geodesic does not reach the origin
(if $\ell\neq 0$). 
We may also consider a special geodesic which stays
at $r=r_0$ for the $p\neq 5$ cases.
For $p=5$, the reality condition for $du/dr$ 
is $\ell^2\le Q_5$, which is independent of the radius.
If $\ell^2 < Q_5$, the geodesic covers the whole 
region of $r$. If $\ell^2=Q_{5}$, the geodesic
stays at a fixed radius (which can take an arbitrary 
value), since we have $dr/du=0$.

We have seen that there are special null geodesics which stay
at a fixed radius, for all the backgrounds in discussion
under certain conditions on $\ell$ and the charge $Q_{p}$
(or $\tilde{Q}_{5}$, or $\tilde{Q}_{1}$). 
We call the point where $dr/du=0$ 
the `critical radius' and denote by $r_0$. 
Precisely, $r_{0}$ is defined as
the point which satisfy
$r^{7-p}+Q_{p}-\ell^{2} r^{5-p}=0$
for the case of $p$-brane solutions; 
and as the point which satisfy
$Q_{p}-\ell^{2} r^{5-p}=0$ when we are considering
the near-horizon geometry of $p$-brane solutions.
($r_0$ means either $r_{0+}$ or $r_{0-}$ when 
there are two critical radii.)
If we take the Penrose limit along such a 
geodesic with a fixed radius,
the geometry becomes independent of $x^+$.
In other words, the resulting background
is the Cahen-Wallach space \cite{CaWa}.

We will present explicit forms of the geometries
in the Penrose limit along a null geodesic at
the critical radius for each brane solution.
For the D$p$-branes, $m_i$ in (\ref{eq:dpmass}) 
at the critical radius can be written in the form
\begin{eqnarray}
&& m_{x}^{2}= {(7-p)Q_{p}\{ (p-7)Q_{p}+2\ell^2 r_{0}^{5-p}\} \over 
8\ell^4 r_{0}^{12-2p} },\nonumber\\
&& m_{y}^{2}= {(7-p)Q_{p}\{ (7-p)Q_{p}-6\ell^2 r_{0}^{5-p}\} \over 
8\ell^4 r_{0}^{12-2p} }, \nonumber\\
&& m_{z}^2={(7-p)Q_{p} \{ (-35+5p)Q_{p}+(30-4p)\ell^2 r_{0}^{5-p} \}\over
 8\ell^4 r_{0}^{12-2p}}.
\end{eqnarray}
In the near-horizon limit, evaluating $m_i$ in 
(\ref{eq:nhDpmass}) with the values of 
$r_0$ for $p\neq 5$ given in (\ref{eq:nhr0}), 
we have
\begin{eqnarray}
 m_{x}^2= m_{z}^2= {(7-p)(p-5) \over 8 \left(Q_{p}\over\ell^2\right)
^{2\over 5-p}}, \quad
m_{y}^2= {(7-p)(1-p)\over 8 \left(Q_{p}\over\ell^2\right)
^{2\over 5-p}}.
\end{eqnarray}
For the D$5$-brane solution, $dr/du=0$ if $\ell^2=Q_{5}$.
In this case, $m_i$ become 
\begin{eqnarray}
 m_{x}^2= m_{z}^2=0, \quad
 m_{y}^2= -{1\over \tilde{r}_{0}^2}
\end{eqnarray}
where $\tilde{r}_{0}$ is a constant which can take
an arbitrary value.

Next we will discuss the NS$5$-branes.
The critical radius is 
$r_0=\sqrt{\ell^2-Q_{5}}$ and 
at this radius, $m_i$ and $H_3$ are evaluated as
\begin{eqnarray}
&&m_{x}^{2}=0, \quad
m_{y}^2= {Q_{5}(Q_{5}-2\ell^2)\over \ell^6},\quad
m_{z}^{2}= {2 Q_{5}(\ell^2-Q_{5})\over \ell^6},\nonumber \\
&&H_3=\frac{2Q_5}{\ell^3}dx^+\wedge dy^1\wedge dy^2.
\end{eqnarray}
In the near horizon limit, 
$dr/du=0$ when $Q_5=\ell^2$ as in the D$5$-brane case. 
In this case, we have $m_x^2=m_z^2=0$, $m_y^2=-1/\ell^2$
and $H_3=\frac{2}{\ell}dx^+\wedge dy^1\wedge dy^2$.

For the fundamental strings, $m_i$ at the critical
radius are given as  
\begin{eqnarray}
&& m_{x}^2= -{3 Q_{1}(3 Q_{1}-\ell^2 r_0^4)\over \ell^2 r_0^{12}},\quad
m_{y}^2= -{3 Q_{1}\over r_0^8},\quad
m_{z}^2= - {3Q_{1}(9Q_1-7\ell^2r_0^4)\over \ell^2r_0^{12}}.
\end{eqnarray}
In the near horizon limit, evaluating 
$m_i$ at the critical radius
$r_0=(Q_1/\ell^2)^{1/4}$, we obtain
$m_x^2=m_z^2=-6\ell^4/Q_1$ and $m_y^2=-3\ell^4/Q_1$.
The NS-NS $3$-form field strength becomes 
$F_3=\frac{6\ell^2}{Q_1^{1/2}}dx^+\wedge 
dx\wedge dz$.

For the $(p,q)$ fivebrane solution, $m_i$
at the critical radius $r_0=\sqrt{\ell^2-\tilde{Q}_{5}}$
become
\begin{eqnarray}
&&m_x^2=\frac{(\ell^2-\tilde{Q}_5)^2\tilde{Q}_5\sin^2\gamma}
{2\ell^4(\ell^2-\tilde{Q}_5\sin^2\gamma)^2}, \nonumber \\
&&m_y^2=-\frac{\tilde{Q}_5(7\ell^4-6\ell^2\tilde{Q}_5+\tilde{Q}_5^2
+(\ell^4+2\ell^2\tilde{Q}_5-\tilde{Q}_5^2)\cos 2\gamma)}
{4\ell^4(\ell^2-\tilde{Q}_5\sin^2\gamma)^2}, \nonumber \\
&&m_z^2=-\frac{(\ell^2-\tilde{Q}_5)\tilde{Q}_5(-9\ell^2+5\tilde{Q}_5
+(\ell^2-5\tilde{Q}_5)\cos 2\gamma)}
{4\ell^4(\ell^2-\tilde{Q}_5\sin^2\gamma)^2}.
\end{eqnarray}
In the near-horizon limit, $dr/du=0$ when
$\tilde{Q}_5=\ell^2$. 
The resulting geometry is given by
\begin{eqnarray}
&& m_{x}^2=m_{z}^2=0,\quad
m_{y}^2=-{1\over \tilde{Q}_5\cos^2\gamma+\tilde{r}_0^2\sin^2\gamma}, 
\nonumber \\
&&H_3=\frac{2\sqrt{\tilde{Q}_5}\cos\gamma}
{\tilde{Q}_5\cos^2\gamma +\tilde{r}_0^2\sin^2\gamma}
dx^{+}\wedge dy^1\wedge dy^2
\end{eqnarray}
where $\tilde{r}_0$ is an arbitrary constant.
This case was studied in ref.~\cite{OzSa}. 
If we choose $\tilde{r}_0^2=\tilde{Q}_5$,
the above background agree with the one given
in ref.~\cite{OzSa}.

In the case of the $(p,q)$ string solution, 
$m_i$  at the critical radius become
 \begin{eqnarray}
   m_{x}^2&=& 
 -{3\tilde{Q}_{1} r_{0}^4
 \over 2 \ell^2 (r_{0}^6+\tilde{Q}_{1}\sin^2\gamma)^3 }
 \Bigl(
 -\tilde{Q}_{1} \ell^2 \cos^2\gamma (-9+7\cos^2\gamma)
 \nonumber\\
 &&
 +2 (3\tilde{Q}_{1}\cos^4\gamma+\ell^6\sin^4\gamma)r_{0}^2
 -\ell^4(3-3\cos^2\gamma+2\cos^4\gamma)r_{0}^4
 \Bigr), \nonumber\\
 m_{y}^2&=&
 {3 \tilde{Q}_{1}(-3+\cos^2\gamma)r_{0}^4\over 2 (
 r_{0}^6+\tilde{Q}_{1}\sin^2\gamma
 )^2},\nonumber\\
 m_{z}^2&=& 
 -{3\tilde{Q}_{1} r_{0}^4
 \over 2 \ell^2 (r_{0}^6+\tilde{Q}_{1}\sin^2\gamma)^3 }
 \Bigl(
 -\tilde{Q}_{1} \ell^2 \cos^2\gamma (-33+19\cos^2\gamma)
 \nonumber\\
 &&
 +2(9\tilde{Q}_{1}\cos^4\gamma+\ell^6\sin^4\gamma)r_{0}^2
 -\ell^4(15-3\cos^2\gamma+2\cos^4\gamma)r_{0}^4
 \Bigr).
 \end{eqnarray}
In the near horizon limit, 
the critical radius is $r_0=(\tilde{Q}_1/\ell^2)^{1/4}$ and
$m_i$ are given by
\begin{eqnarray}
 m_{x}^2&=& m_{z}^2=
-{3\tilde{Q}_{1}^3
\left(
2\tilde{Q}_{1}\cos^4\gamma+3\ell^4 r_{0}^2 \cos^2\gamma\sin^2\gamma
+\ell^6 \sin^4\gamma
\right)
\over \ell^6 r_{0}^2 
(r_{0}^6\cos^2\gamma+\tilde{Q}_{1}\sin^2\gamma)^3},\nonumber
\\
m_{y}^2&=&
-{3\tilde{Q}_{1}^3\cos^2\gamma
\left(
\tilde{Q}_{1}\cos^2\gamma+\ell^4r_{0}^2\sin^2\gamma
\right)
\over \ell^6 r_{0}^2 
(r_{0}^6\cos^2\gamma+\tilde{Q}_{1}\sin^2\gamma)^3}.
\end{eqnarray}

\section{Strings on the Penrose limits of various branes}
Having obtained the Penrose limits of the various
brane solutions, we shall now study string theories 
on these backgrounds.
In the present paper, we concentrate on the analysis of 
the bosonic sector. We plan to report on the analysis 
of the fermionic sector and on the related issues such as
the condition on the background
for the realization of the world-sheet 
supersymmetry in the light-cone gauge in a future work.
In section 4.1, we give a brief review on string 
theory on the plane wave background. We derive the
equations of motion for the bosonic string in the 
light-cone gauge, and see that we have massive theories
with time-dependent masses in general.
Then in section 4.2, we start the study of
the equations of motion for each case.  
We discuss the structures of the singular points of 
the differential equations, especially.
In section 4.3, we point out that in 
certain circumstances, the equations of motion 
generically take simple forms. The first example
is the theory on the critical radius, 
for which the masses become constant. Another example
is the near-horizon limit.
In this limit, the equations of motion for a large class
of backgrounds can be solved by the Bessel functions.
The case of the fivebranes, for which the exact 
solutions exist, will be discussed separately in detail
in section 5.

\subsection{Light-cone string theory on the plane wave background}
The sigma-model action for the bosonic part of string theory
is given by
\begin{equation}
S={1\over 4\pi \alpha'}\int\! d^2\sigma \left(\sqrt{-h}
h^{\alpha\beta}g_{\mu\nu}
\partial_\alpha X^\mu\partial_\beta X^\nu
+\epsilon^{\alpha\beta}B_{\mu\nu}
\partial_\alpha X^\mu\partial_\beta X^\nu\right)
\label{sigmamodel}
\end{equation}
where $h_{\alpha\beta}$ ($\alpha,\beta=\tau,\sigma$) is the 
world-sheet metric and $\epsilon^{\tau\sigma}=+1$.
$g_{\mu\nu}$, $B_{\mu\nu}$ ($\mu,\nu=0,\ldots,9$) are the 
background metric and NS-NS two-form potential, respectively.
The RR backgrounds do not couple to the bosonic sector.

The plane wave metrics obtained in section 2 are 
of the form (\ref{Brinkman}):
\[
ds^2=2dx^{+}dx^{-}+
\left(m^{2}_{x}(x^{+}) x_{a}^{2}+m_{y}^{2} (x^{+}) y_{l}^2
+m_{z}^2(x^{+})z^2\right)
(dx^{+})^2+ dx_{a}^{2}+ dy_{l}^{2}+ dz^2.
\]
The directions represented by $x_a$ $(a=1,\ldots, p)$
are the ones which were originally the spatial
directions of the world-volume of the $p$-branes,
and those represented by $y_l$ $(l=1,\ldots, 7-p)$ 
come from the angular coordinates of the transverse
$(7-p)$-sphere. The coordinate
$z$ comes from the angular coordinate which appears
in (\ref{trp metric}). In the following,
we frequently denote those $x$-, $y$- and $z$- directions
collectively as $x^i$ $(i=1,\ldots,8)$. 

For the fundamental and $(p,q)$ string and the 
NS and $(p,q)$ fivebrane solutions, 
there are non-vanishing NS-NS $B$-fields.
As we have seen, the only non-vanishing component of the
field strength is $F_{+ij}$. We take the non-vanishing 
component of the $B$-field in the form
\[
B_{ij}=B_{ij}(x^+)
\]
by choosing the gauge.

We adopt the conformal gauge ($\sqrt{-h}h^{\alpha\beta}=
\eta^{\alpha\beta}$) for the world-sheet
metric, and substitute the plane wave background 
into the action. Then, (\ref{sigmamodel}) becomes
\begin{eqnarray}
S&=&{1\over 4\pi \alpha'}\int\! d^2\sigma \Big(
2\partial_{\alpha} X^{+}\partial^{\alpha} X^{-} 
+ \partial_{\alpha} X^{i}\partial^{\alpha} X^{i}
\nonumber\\
&&\hspace*{2cm}
+m^{2}_{i}(X^{+})\partial_{\alpha} X^{+}\partial^{\alpha} 
X^{+} (X^{i})^2
+\epsilon^{\alpha\beta}B_{ij}(X^{+})\partial_{\alpha} 
X^{i}\partial_{\beta} X^{j}\Big).
\label{ppsigmamodel}
\end{eqnarray}
We can take the light-cone gauge $X^{+}=\tau$,
since it is consistent with the equation
of motion derived from (\ref{ppsigmamodel}) \cite{HoSte}.
In the light-cone gauge, the fields in the transverse
directions $X^{i}$ are effectively described by the action
\[
S={1\over 4\pi \alpha'}\int\! d^2\sigma \left(
-\partial_{\tau} X^{i}\partial_{\tau} X^{i} 
+\partial_{\sigma} X^{i}\partial_{\sigma} X^{i}
-m^{2}_{i}(\tau)(X^{i})^{2}
+\epsilon^{\alpha\beta}B_{ij}(\tau)
\partial_{\alpha} X^{i}\partial_{\beta} X^{j}
\right),
\]
which gives the equation of motion 
\begin{equation}
\partial^{2}_{\tau} X^{i} 
-\partial^{2}_{\sigma} X^{i} -m_{i}^{2}(\tau) X^{i} 
-\sum_{j}\partial_{\tau} B_{ij}(\tau)
\partial_{\sigma} X^{j}=0.
\label{ppeom}
\end{equation}
We have obtained a massive theory whose mass is dependent
on the world-sheet time. Note that in our sign convention,
$m^{2}_{i}<0$ is massive and $m^{2}_{i}>0$ is tachyonic.

Expanding $X^{i}$ in the Fourier components with respect to 
$\sigma$ ($0\le \sigma \le 2\pi$) 
\begin{equation}
X^{i}(\tau,\sigma)=\sum_{n=-\infty}^{\infty}\left(\alpha_{n}^{i}
\varphi^{i}_{n}(\tau)+ \tilde{\alpha}_{-n}^{i}
\tilde{\varphi}^{i}_{-n}(\tau) \right)
e^{in\sigma},
\end{equation}
we obtain the equations of motion for the functions 
$\varphi^{i}_{n}$:
\begin{equation}
{d^2\over d\tau^2}\varphi^{i}_{n}
+n^2\varphi^{i}_{n} -m^{2}_{i}(\tau)\varphi^{i}_{n}
-in\sum_{j}\partial_\tau B_{ij}(\tau)\varphi^{j}_{n} =0.
\label{modeeom1}
\end{equation}
$\tilde{\varphi}^{i}_{-n}$ satisfy the same equations
of motion.

In the remaining part of the paper, 
we analyze the equation of motion 
(\ref{modeeom1}) for the plane wave background for each 
brane solution. Though the main aim of our paper
is to find solutions of the equation, we  
would like to make a few comments on the
quantization of the string here.
By imposing the canonical commutation relations
on the fields $X^i$ and the conjugate momenta, 
we obtain the oscillator commutation relations for 
$\alpha_{n}^{i}$ and $\tilde{\alpha}_{n}^{i}$, 
{\it provided} that the functions
$\varphi_{n}$ and $\tilde{\varphi}_{n}$ are the solutions
of the equation of motion. In contrast to the string theory
on the flat background, the light-cone Hamiltonian
does not commute with the number operator, for the case
of the time-dependent plane wave background.
This situation will be illustrated in the appendix when 
we give an example of the mode expansion
using Hankel functions as a basis.

We should also comment on the dilaton coupling. That
coupling  is not taken into account
in this paper, that is,  we study the propagation of a free
string. When we are considering the brane solutions which
have finite dilaton background throughout the spacetime,
and if we set the dilaton v.e.v. at infinity to 
a small value, this gives a good description.
However, when we discuss the branes with the
dilaton which diverges at the origin (the D$p$-branes with
$p\le 2$, the $(p,q)$ fivebranes other than the D5-branes,
and $(p,q)$ strings other than the fundamental strings),
care is needed.
If we consider a geodesic which passes the origin in 
such backgrounds, the effect of the interaction 
will become important.

\subsection{Equations of motion for each brane solution}
The equations of motion of the string on the Penrose limits
of the brane solutions are given by substituting 
$m^{2}_{i}$ and $B_{ij}$ obtained in section 2 
into the general form (\ref{modeeom1}).
\medskip

\noindent
{\bf (1) D$p$-branes}
\medskip

\noindent
We first discuss the case of the D$p$-branes.
We have $B_{ij}=0$, and the equation
of motion is given by
\begin{equation}
\left({d^2\over d\tau^2}+n^{2}-m_{i}^2(\tau) \right)\varphi_{n}=0.
\label{DE1}
\end{equation}
The masses  $m^{2}_{i}(\tau)$
are functions of $\tau$ through the radial coordinate $r(\tau)$.
The relation between $r$ and $\tau$ is given in the
differential form as 
\begin{equation} 
{d\tau\over dr}\equiv Q(r)=\left({r^{7-p}+Q_{p}\over 
r^{7-p}+Q_{p}-\ell^2 r^{5-p}}\right)^{1\over2}.
\label{eq:QrDp}
\end{equation}
Note that this is obtained from (\ref{QDp}) by
simply replacing $u$ with $\tau$, because we are working 
in the light-cone gauge $\tau=x^{+}=u$.

To find solutions of the differential equation,
it is convenient to take $r$ as an independent variable 
instead of $\tau$. The equation
(\ref{DE1}) is rewritten as
\begin{equation}
\left({d^2\over dr^2}-{1\over Q(r)}{dQ(r)\over dr} {d\over dr}
+(n^2-m_{i}^2(r))Q^2(r)\right) \varphi_{n}=0.
\label{DE2}
\end{equation}
For the case of the D$p$-branes with odd $p$ ($p=2k+1$ where $k=0,1,2$),
we will use $x=r^2$ as the independent variable. 
The equation of motion is transformed into 
\begin{equation}
\left({d^2\over dx^2} +p(x){d\over dx}+q(x)\right) \varphi_{n}=0
\label{DEoddp}
\end{equation}
where 
\begin{eqnarray}
p(x)&=& {1\over 2x}-{1\over Q}{d Q\over dx},\label{eq:defpx}\\
q(x)&=& {1\over 4x}(n^2-m_{i}^2)Q^2\label{eq:defqx}.
\end{eqnarray}

Let us examine the structure of the singular points
of the differential equation for the odd $p$ case (\ref{DEoddp}). 
A singular point $x_{0}$ is the (complex) value of 
$x$ where $p(x)$ or $q(x)$ diverge as $x\rightarrow x_{0}$. 
If $p(x)=O((x-x_{0})^{-1})$ and $q(x)=O((x-x_{0})^{-2})$, 
$x_{0}$ is called a regular singular point. 
If this is not the case, $x_{0}$ is
called an irregular singular point. 
For our case, the coefficient $p(x)$ is
obtained using (\ref{eq:QrDp}) as
\begin{equation}
p(x)={1\over 2x}-{(3-k) x^{2-k}\over 2(x^{3-k}+Q_p)}
+{(3-k)x^{2-k}-(2-k)\ell^2 x^{1-k}\over
2(x^{3-k}+Q_p-\ell^2 x^{2-k})}.
\label{pxDp}
\end{equation}
The coefficient $q(x)$ for the cases of the
$x$-, $y$- or $z$-directions are calculated 
by substituting  $m^2_{x}$, $m^2_{y}$ or $m^2_{z}$ 
in (\ref{eq:defqx}), respectively.

For the D$5$-branes ($k=2$), (\ref{DEoddp}) has four 
singular points.
As we can see from the denominators of $p(x)$ and $q(x)$,
we have regular singular points at $x=0,Q_p, \ell^2-Q_p$. 
There is an irregular singular point at $x=\infty$,
which can be seen by transforming  the variable $x$ to $1/x$.
As $k$ decreases, the number of the singular points increases.
For the D$3$-branes ($k=1$), we have regular 
singular points at $x=0,\pm i \sqrt{Q_p}, 
(\ell^2\pm\sqrt{\ell^2-4Q_p})/2$.
For the D$1$-branes ($k=0$), we have regular 
singular points at $0$, the three roots of $x^3+Q_p=0$, 
and the three roots of $x^3+Q_p-\ell^2x^2=0$.
The irregular singular point at $x=\infty$ is present for 
all the cases.

For D$p$-branes with even $p$, use of the above $x$ does
not lead to rational expressions, and we use the 
variable $r$ and examine the differential equation
(\ref{DE2}). The coefficient of $d\varphi_{n}/dr$ is
given by
\begin{equation}
-{1\over Q(r)}{dQ(r)\over dr}=
-{(7-p) r^{6-p}\over 2(r^{7-p}+Q_p)}
+{(7-p)r^{6-p}-(5-p)\ell^2 r^{4-p}\over
2(r^{7-p}+Q_p-\ell^2 x^{5-p})}.
\end{equation}
Generically, there are regular
singular points at $0$, the roots of $r^{7-p}+Q_{p}=0$,
and the roots of $r^{7-p}+Q_p-\ell^2 r^{5-p}=0$, and
an irregular singular point at $r=\infty$. 

In general, differential equations having fewer number
of singular points are more tractable. (See e.g. 
ref.\cite{In}.) For the differential equations 
for the D5-branes, we can find non-trivial solutions 
as we see in section 5.
It should also be noted that it may be possible to
reduce the number of the singular points 
by the change of variables or by factoring out
some function from $\varphi_{n}$.
Indeed, the equation in the $y$-directions for the  
D5-branes can be transformed to the one which has only 
two regular singular points at $x=0, \ell^2-Q_p$ and
an irregular singular point at $x=\infty$, as we discuss
in section 5. 

We comment here that the differential equations 
in the $y$- and $z$-directions for the D$6$-branes
in the $\ell=0$ case can be transformed into the
form which is found in the literature (eq.~VII, p.~503 
of ref.~\cite{In}). The properties of the solution 
of that equation have been  investigated to some 
extent \cite{In2}, but we do not discuss
this case any further in this paper.
\medskip

\noindent
{\bf (2) NS5-branes and $(p,q)$ fivebranes}
\medskip

\noindent
In these cases, we have non-vanishing NS-NS $B$-fields 
in the two $y$-directions. Let us consider the NS5-brane case. 
The equations of motion (\ref{ppeom}) for the $y$-directions
are coupled equations 
\begin{eqnarray}
\partial^{2}_{\tau} \varphi_{n}^{1} +n^{2} \varphi_{n}^{1}
-m_{y}^{2}(\tau) \varphi_{n}^{1} -i n \partial_{\tau} B(\tau)
\varphi_{n}^{2}=0,\nonumber\\
\partial^{2}_{\tau} \varphi_{n}^{2} +n^{2} \varphi_{n}^{2}
-m_{y}^{2}(\tau) \varphi_{n}^{2} +i n \partial_{\tau} B(\tau)
\varphi_{n}^{1}=0,
\label{eq:eomB}
\end{eqnarray}
where $B(\tau)\equiv B_{12}=-B_{21}$ and the indices
$1,2$ denote the two $y$-directions.
Explicit form of $\partial_\tau B$ for  the
NS5-branes is given by
\begin{equation}
\partial_{\tau} B(\tau)= \partial_{x^{+}} B_{12}(x^{+})=H_{+12}
= {2 \ell Q_{5}\over (r^{2}+Q_{5})^{2}}.
\label{eq:dtB}
\end{equation}
Note that we have used the light-cone gauge condition 
$\tau=x^{+}$ in the first equality.
The equation (\ref{eq:eomB}) is diagonalized by the combination 
$\varphi_{n}^{(\pm)}=\varphi_{n}^{1}\mp i \varphi_{n}^{2}$:
\begin{equation}
\Big( \partial^{2}_{\tau} +n^{2} -m^{2}_{y}(\tau)
\pm n \partial_{\tau} B(\tau) \Big) \varphi_{n}^{(\pm)}=0.
\label{DEwithB}
\end{equation}

We use $x$ as the independent variable and analyze the
differential equation of the form (\ref{DEoddp}). 
The coefficient $p(x)$ which is given by the formula
(\ref{eq:defpx}) is evaluated as
\begin{equation}
p(x)={1\over x}+{-1\over x+Q_{5} }
+{{1\over2}\over x+Q_{5}-\ell^2}
\label{pxNS5}
\end{equation}
using $Q(r)$ obtained in (\ref{QNS5}).
The formulae for $q(x)$ for $\varphi_{n}^{(\pm)}$ 
are modified to 
\begin{equation}
q(x)= {1\over 4x}(n^2-m_{y}^2\pm \partial_{\tau} B)Q^2.
\end{equation}
It is evaluated using $m^{2}_{y}$ and $\partial_{\tau} B$ given
in (\ref{eq:NS5mass}) and (\ref{eq:dtB}), respectively.
As for the $x$- and $z$-directions, the differential equation
is given by (\ref{DE1}). By using $x$, we obtain the differential
equation with $p(x)$  given by (\ref{pxNS5}),
$q(x)$  given by (\ref{eq:defqx}) with the substitution of
$m^{2}_{x}$ and $m^{2}_{z}$ which are
found in (\ref{eq:NS5mass}). 
Note that the equation of motion for 
$x$-directions is the one for the flat background,
since $m^{2}_{x}=0$. 
The differential equations have three regular singular
points at $x=0, -Q_{5}, \ell^2-Q_{5}$, and an irregular
singular point at $x=\infty$. 

The $(p,q)$ fivebrane case can be analyzed exactly in the
same way as for the NS5-brane case. We have the coupled
equations (\ref{eq:eomB}) where
\begin{equation}
\partial_\tau B(\tau)= {2\tilde{Q}_5\ell\cos\gamma \over 
(r^2+\tilde{Q}_5)(r^2+\tilde{Q}_5 \cos^2\gamma)}.
\label{eq:tBpq5}
\end{equation}
The coefficient $p(x)$ is evaluated using $Q(r)$ in
(\ref{Q(p,q)5}) and is given by
\begin{equation}
p(x)={1\over x}+{-{1\over 2}\over x+\tilde{Q}_{5} }
+{{1\over2}\over x+\tilde{Q}_{5}-\ell^2}
+{-{1\over 2}\over x+\cos^2\gamma \tilde{Q}_{5} }.
\end{equation}
The differential equations have four regular singular
points at $x=0, -\tilde{Q}_{5}, \ell^2-\tilde{Q}_{5}, 
-\cos^2\gamma \tilde{Q}_{5}$, 
and an irregular singular point at $x=\infty$. 
The differential equations for the $y$-directions 
for the NS5-branes (or the $(p,q)$ fivebranes) 
can be transformed to the ones which
have only two regular singular points $x=0, \ell^2-Q_{5}$
(or $\ell^2-\tilde{Q}_{5}$) and an irregular singular 
point at $x=\infty$, as in the case of the D5-branes.
\medskip

\noindent
{\bf (3) fundamental strings and $(p,q)$ strings}
\medskip

\noindent
The non-vanishing component of the $B$-field is
$B_{xz}$ in these cases. The equations of motion
in the $x$- and $z$-directions are coupled equations
\begin{eqnarray}
\partial^2_\tau \varphi_{n}^x +n^2 \varphi_{n}^x
-m_x^2(\tau) \varphi_{n}^x -i n \partial_\tau B(\tau)
\varphi_{n}^z=0,\nonumber\\
\partial^2_\tau \varphi_{n}^z +n^2 \varphi_{n}^z
-m_z^2(\tau) \varphi_{n}^z +i n \partial_\tau B(\tau)
\varphi_{n}^x=0,
\label{eq:eomBF1}
\end{eqnarray}
where $B(\tau)\equiv B_{xz}=-B_{zx}$. 
Note that $m^2_x\neq m^2_z$ for the present cases.

The $B$-field for the fundamental strings is given by
\begin{equation}
\partial_\tau B(\tau)= \partial_{+} B_{xz}(x^+)=F_{+xz}
= {6 \ell Q_1\over r^2 (r^6+Q_1)}
\end{equation}
and the masses  $m^{2}_{i}$ are given in 
(\ref{eq:F1mass}).
Diagonalizing (\ref{eq:eomBF1}), the equations of
motion in the $x$- and $z$-directions become
\begin{equation}
\Big( \partial^2_\tau +n^2 -\bar{m}^2(\tau)
\pm \Delta m^2 \Big) \varphi_{n}^{(\pm)}=0
\label{eq:F1diag}
\end{equation}
where 
\begin{eqnarray}
\bar{m}^2&\equiv&{m_x^2+m_z^2\over 2}
= -3 {Q_{1}(Q_{1}^2+Q_{1}\ell^2 r^4+5Q_{1}r^6-8\ell^2 r^{10}+
4r^{12})
\over r^8 (r^6+Q_{1})^2},  \\
\Delta m^2&\equiv& {3 Q_1\over r^2 (r^6+Q_1)}\sqrt{4n^2\ell^2+9}.
\end{eqnarray}
The differential equations can be analyzed using 
$x$ as the independent variable. Number of 
the singular points is the same as the one for the
D1-brane case.

The $B$-field for the $(p,q)$ strings is given by
\begin{equation}
\partial_\tau B(\tau)= 
{6\ell \tilde{Q}_1 \cos\gamma r^4 \over 
(r^6+\tilde{Q}_1)(r^6+\sin^2\gamma \tilde{Q}_1)}
\end{equation}
and the masses are given in (\ref{pq1mz}).
Diagonalization of the coupled equations is
performed as in the case of the fundamental
string and gives (\ref{eq:F1diag}) with
\begin{eqnarray}
\bar{m}^2&\equiv&{m_x^2+m_z^2\over 2}, \\
\Delta m&\equiv& {3 \tilde{Q}_1 r^4\over (r^6+\tilde{Q}_1) 
(r^6+\sin^2\gamma \tilde{Q}_1)}
\sqrt{4n^2\ell^2\cos^2\gamma+9}.
\end{eqnarray}
The differential equations for the $(p,q)$ strings have
extra singular points at the roots of $x^3+\sin^2\gamma 
\tilde{Q}_{1}=0$, compared to the case of the fundamental
strings.

\subsection{Analysis of the simply solvable theories}
As we mentioned in section 3, it is possible to consider 
a geodesic for which $r$ is kept on a fixed radius $r=r_{0}$.
The Penrose limit along such a geodesic gives the string theory
with constant mass. 
The equation of motion (for the $B_{ij}=0$ case)
\begin{equation}
\left({d^2\over d\tau^2}+n^2-m_{i}^2(r_{0}) \right)\varphi_{n}=0
\end{equation}
is easily solved by 
\begin{eqnarray}
&&\varphi_{n}=e^{i\omega_{n}\tau},\qquad
\tilde{\varphi}_{-n}=e^{-i\omega_{n}\tau},\nonumber\\
&&\omega_{n}=\sqrt{n^2-m_{i}^2(r_{0})}.
\end{eqnarray}
When $B_{ij}\neq 0$, we can start from the diagonalized 
equation of the forms (\ref{DEwithB}) and (\ref{eq:eomBF1})
and find similar solutions.

Quantization of the strings should be performed following
the standard procedure. (See e.g. ref.\cite{MeTs}.)
We note here that there are some cases for which 
$m^{2}_{i}(r_{0})>0$. It seems that tachyonic states
are present in the resulting theories.
Whether we can construct 
physically meaningful string theories for such cases
is not clear to us at present.\footnote{Some discussions 
on the tachyonic mass from the standpoint 
of the holographic renormalization group  
are given in ref.\cite{GiPaSo}.}

In the near-horizon limit
\begin{equation}
r\ll Q_{p}^{1\over 7-p},
\label{nhlimit}
\end{equation}
the string theory also becomes simple.
The masses in the near-horizon limit for the 
case of the D$p$-branes are given in (\ref{eq:nhDpmass}),
while the relation between $\tau$ and $r$ becomes
\begin{equation}
{d\tau\over dr}=
\left({Q_{p}\over Q_{p}-\ell^2 r^{5-p}}\right)^{1\over2}.
\label{nhQ}
\end{equation}
Firstly, for the case of D3-branes, 
the masses become constant, as mentioned in
section 2.
Next, for $p=5$, (\ref{nhQ}) states that $r$ is
proportional to $\tau$, and the masses  become 
\begin{eqnarray}
&&m_{x}^2=m_{z}^2={1\over 4\tau^{2}},\nonumber\\
&&m_{y}^2=-{3\ell^2+Q_{p}\over Q_{p}-\ell^2}{1\over \tau^2}.
\label{eq:nhD5mass}
\end{eqnarray}
The equations of motion take the form 
\begin{equation}
\left({d^2\over d\tau^2}+n^2
- {C_{1}\over \tau^{2}} \right)\varphi_{n}=0
\label{eq:eomBessel1}
\end{equation}
where $C_{1}$ is a constant which can be read off from 
(\ref{eq:nhD5mass}). The equation (\ref{eq:eomBessel1}) 
can be transformed to the Bessel equation (when $n\neq 0$)
\begin{equation}
\left({d^2\over d\tau'{}^2}+{1\over \tau'}{d\over d\tau'}
+1- {\nu^2\over \tau'{}^{2}} \right)
\tau'{}^{-{1\over 2}}\varphi_{n}=0
\end{equation}
where $\tau'\equiv n\tau$ and $\nu^2\equiv C_{1}+1/4$.

For the cases $p=0,1,2,4$, 
we take the limit
\begin{equation}
r\ll \left(Q_{p}\over \ell^2\right)^{1\over 5-p}\qquad(p=0,1,2,4),
\label{nhlimit2}
\end{equation}
in addition to the near-horizon limit.
Then, the second terms in the denominator of (\ref{eq:nhDpmass})
can be neglected, and the relation between $\tau$ and $r$ 
(\ref{nhQ}) become $\tau=r$. Note that if we take the radial 
null geodesic ($\ell=0$), (\ref{nhlimit2}) is always satisfied.
In this limit, the masses become
\begin{equation}
m_{x}^2=m_{y}^2=m_{z}^2={(7-p)(3-p)\over 16 \tau^2}
\end{equation}
and we obtain the equations of motion 
of the form (\ref{eq:eomBessel1}).

In the appendix, we present a preliminary study for the canonical
quantization of the string on this kind of plane wave background.
Using the solutions of the equation of motion as a basis,
we construct oscillators and obtain the light-cone Hamiltonian.

\section{Solutions for the fivebrane backgrounds}
In this section, we study string theories on the 
Penrose limit of the D5-brane, NS5-brane and $(p,q)$ 
fivebrane solutions. 
These backgrounds are of particular interest, because 
we can solve the equations of motion of strings 
in the two regions: near the critical radius, and
far from the origin.
We study the former limit in section 5.1 and
the latter limit in section 5.2.
Furthermore, for a particular component,
the equations of motion can be solved exactly.
In section 5.3, we describe the exact solutions
which are written
by a special function called the spheroidal wave function.

\subsection{Solutions near the critical radius}
We firstly consider the D5-brane case.
As discussed in section 4, the equation of motion of the
string leads to the Schr{\"o}dinger type equation of the form
\begin{equation}
\left({d^2\over d\tau^2}+n^2-m_{i}(\tau)^2 \right) 
\varphi_{n}(\tau)=0.
\label{eq:schroe}
\end{equation}
Explicit forms of $m_{i}^2$ are given by 
the $p=5$ case of (\ref{eq:dpmass}):
\begin{eqnarray}
 m_{x}^{2}&=&-{1\over4}{Q_{5}(Q_{5}^2-Q_{5}\ell^2
+(7Q_{5}-8\ell^2)r^2+6r^4)\over r^2(r^2+Q_{5})^3},
\label{eq:mxD5}\\
m_{y}^{2}&=&-{1\over4}{Q_{5}(Q_{5}^2+3Q_{5}\ell^2
+(-Q_{5}+8\ell^2)r^2-2r^4)\over r^2(r^2+Q_{5})^3},
\label{eq:myD5}\\
m_{z}^{2}&=&-{1\over4}{Q_{5}(Q_{5}^2-Q_{5}\ell^2
-(Q_{5}+8\ell^2)r^2-2r^4)\over r^2(r^2+Q_{5})^3}.
\label{eq:mzD5}
\end{eqnarray}
The relation between $\tau$ and $r$ is given from
(\ref{QDp}) as
\begin{equation}
{d\tau\over dr}=Q(r)
=\left({r^2+Q_{5}\over r^2+Q_{5}-\ell^2}\right)^{1\over2}.
\label{eq:QrD5}
\end{equation}

As we have seen in section 3, if $\ell^2>Q_{5}$,
the geodesic in the fivebrane background stops at a critical 
radius $r_{0}=\sqrt{\ell^2-Q_{5}}$. 
We consider this kind of geodesics here, and
study the theory near $r=r_{0}$. 

Taking the leading term in the expansion of (\ref{eq:QrD5})
around $r=r_{0}$, we note that the relation between 
$\tau$ and $r$ becomes
\begin{eqnarray}
&&\tau =\int dr {\ell\over \sqrt{r^2-r_{0}^2}}
=\ell\, {\rm arccosh} {r\over r_{0}},\nonumber\\
&&r=r_{0}\,{\rm cosh }{\tau\over \ell}.
\label{eq:trD5CR}
\end{eqnarray}
Now we expand the masses around $r=r_{0}$ and take the 
leading correction to the constant mass
\begin{equation}
m^{2}_{i}(r)=m^{2}_{i}(r_0)+{\partial m^{2}_{i}(r_{0})\over 
\partial r^{2}} (r^2-r_{0}^2).
\label{eq:massexpand}
\end{equation}
Substituting this form of mass term and using the 
relation (\ref{eq:trD5CR}), we find that the equation of motion
(\ref{eq:schroe}) takes the 
form of the (modified) Mathieu equation \cite{Er}
\begin{equation}
\left({d^2\over dz^2}-h+2\theta \,{\rm cosh} 2z\right)\varphi_{n}=0.
\label{eq:mathieu}
\end{equation}
where $z\equiv\tau/\ell$.
The parameters $h$ and $\theta$ are given by 
\begin{eqnarray}
h&=&-\ell^2 \left(n^2 -m^{2}_{i}(r_0)
+{r_{0}^2\over 2}{\partial m^{2}_{i}(r_{0})\over 
\partial r^{2}}\right),\nonumber\\
\theta&=&-{r_{0}^2\ell^2\over 4}{\partial m^{2}_{i}(r_{0})
\over \partial r^{2}}.
\end{eqnarray}
When evaluated using the mass terms
(\ref{eq:mxD5})-(\ref{eq:mzD5}), they become
\begin{eqnarray}
&&h=-\ell^{2} n^{2} +{Q_{5} (16\ell^2-11 Q_{5})\over 
8\ell^4},\quad\hspace{1.5cm} \theta={Q_{5}(12\ell^2-11 Q_{5})\over 16 \ell^4}
\qquad\mbox{($x$-directions)},\nonumber\\
&&h=-\ell^{2} n^{2} -{Q_{5} (32\ell^{4}-29Q_{5}\ell^{2} 
+9 Q_{5}^{2})\over 8\ell^4 r_{0}^2},\quad 
\theta=-{Q_{5}(20\ell^{4}-25Q_{5}\ell^{2}+9 Q_{5}^{2})
\over 16 \ell^4 r_{0}^{2}}\nonumber\\
&&\qquad\hspace*{11.5cm}\mbox{($y$-directions)},\nonumber\\
&&h=-\ell^{2} n^{2} +{3 Q_{5} (16\ell^2-9 Q_{5})\over 
8\ell^4},\quad\hspace{1.5cm} 
\theta={Q_{5}(28\ell^2-27 Q_{5})\over 16 \ell^4}
\qquad\,\,\mbox{($z$-direction)}.\nonumber\\
\end{eqnarray}

Next we consider the NS5-branes. The equations of motion
in the $x$- and $z$-directions are given by (\ref{eq:schroe})
and the ones for the $y$-directions are 
\begin{equation}
\Big( \partial^2_\tau +n^2 -m_{y}(\tau)^2
\pm n \partial_\tau B(\tau) \Big) \varphi^{(\pm)}_{n}=0.
\label{eq:eomB5}
\end{equation} 
The relation between $\tau$ and $r$ is 
\begin{equation}
 {d\tau\over dr}={r^2+Q_{5}\over r\sqrt{r^2+Q_{5}-\ell^2}},
\end{equation}
which leads to 
\begin{equation}
r=r_{0}\,{\rm cosh }{r_{0}\tau\over \ell^{2}}
\label{eq:trNS5CR}
\end{equation}
in the $r\rightarrow r_{0}$ limit.
Expanding the masses and $\partial_{\tau}B$ in 
(\ref{eq:NS5mass}) and (\ref{eq:dtB}) around $r=r_{0}$, 
and substituting them into the equations of motion,
we obtain the modified Mathieu equation (\ref{eq:mathieu})
with parameters:
\begin{eqnarray}
&&h=-{\ell^{4} n^{2}\over r_{0}^2}
-{Q_{5} (6\ell^{4}-8Q_{5}\ell^{2} 
+3 Q_{5}^{2}\mp 2n\ell^3(2\ell^2-Q_{5}))
\over \ell^4 r_{0}^2},\quad 
\theta=-{Q_{5}(4\ell^{2}-3 Q_{5}\mp2n\ell^3)
\over 2 \ell^{4}}\nonumber\\
&&\hspace*{10cm}\mbox{for }\varphi^{(\pm)}_{n}
\mbox{ ($y$-directions),}
\nonumber\\
&&h=-{\ell^{4} n^{2}\over r_{0}^2} 
+{Q_{5} (4\ell^{2}-3 Q_{5})\over \ell^{4}},\quad 
\theta={Q_{5}(2\ell^{2}-3 Q_{5})\over 2 \ell^{4}}
\qquad\hspace{1.5cm}\,\,\mbox{($z$-direction)}
\end{eqnarray}
Since $m^{2}_{x}=0$, the equations in the $x$-directions
are the same as the ones for the flat space.
The equations of motion for the $(p,q)$ fivebrane case
can be brought to the modified Mathieu equation,
following the same steps as above.

The solutions to the equations of motion is given by
the Mathieu functions. 
Since the theory which we are considering is an
approximation around a point in the spacetime,
it is not clear what kind of boundary conditions
should be imposed on the solutions, as it stands.
It would be interesting to study the solutions which
match the ones in the $r\rightarrow\infty$ region which are
given in the next subsection. It should enable us to
investigate the whole region of the geodesic. 
This problem will be discussed elsewhere. 

\subsection{Asymptotic solutions at infinity}
We shall now study the $r\rightarrow\infty$ limit.
First consider the D5-brane example.
As we see from (\ref{eq:dpmass}), the mass terms for all the
directions are proportional to $1/r^4$ asymptotically
\begin{equation}
m^{2}_{i}\rightarrow -{C_{2} \over r^4}
\label{eq:asmass}
\end{equation}
where the constant $C_{2}$ is given by
\begin{equation}
C_{2}={3\over 2}Q_{5}\quad\mbox{($x$-directions)},\qquad 
C_{2}=-{1\over 2}Q_{5}\quad\mbox{($y$- and $z$-directions)}.
\end{equation}
Moreover, we have $\tau=r$, as we can see from 
the $r\rightarrow\infty$ limit of (\ref{QDp}).

By using $x=r^2=\tau^2$ as the independent variable,
the equation of motion (\ref{eq:schroe}) reads
\begin{equation}
\left({d^2\over dx^2} + {1\over 2x}{d\over dx}
+{n^2\over 4x}+ {C_{2} \over 4x^3}\right) \varphi_{n}=0.
\end{equation}
This differential equation has two irregular singular
points at $x=0$ and $x=\infty$. This equation is transformed 
to the standard form \cite{In} by the redefinitions 
$\psi_{n}=x^{-1/4}\varphi_{n}$ and $z=nx/\sqrt{C_{2}}$:
\begin{equation}
\left(z^2 {d^2\over dz^2} + z{d\over dz}
-{1\over 4}\{ h -\theta (z+{1\over z})\}\right)
\psi_{n}=0,
\label{eq:stmathieu}
\end{equation}
where 
\begin{equation}
h={1\over 4},\qquad \theta=\sqrt{C_{2}} n.
\end{equation}
By the substitution $z=e^{2iw}$, (\ref{eq:stmathieu}) becomes
the Mathieu equation \cite{Er}
\begin{equation}
\left({d^2\over dw^2}+h-2\theta \cos 2w\right)\psi_{n}=0.
\label{eq:mathieu2}
\end{equation}

For the cases of the NS5-branes or the $(p,q)$ fivebranes, 
the discussion is essentially the same.
Asymptotic behaviors of the masses $m^{2}_{x}$
and $m^{2}_{z}$ for the $(p,q)$ fivebranes
are given by (\ref{eq:asmass}) with
\begin{equation}
C_{2}={3\over 2}\sin^2\gamma\tilde{Q}_{5}\quad 
\mbox{($x$-directions)},\qquad
C_{2}=-{1\over 2}(1+3\cos^2\gamma)\tilde{Q}_{5}\quad 
\mbox{($z$-directions)}.
\end{equation}
The differential equations for these directions
are brought to the Mathieu equations 
with the $C_{2}$ defined above.

We also note that the  asymptotic forms of $m^{2}_{y}$
and $\partial_{\tau}B$ are
\begin{equation}
m^{2}_{y}\rightarrow {(1+3\cos^2\gamma)\tilde{Q}_{5}\over
2 r^{4}},\qquad
\partial_\tau B\rightarrow {2\ell\cos\gamma\tilde{Q}_{5}
\over r^4}.
\end{equation}
The equations of motion for $\varphi^{(\pm)}_{n}$ in
the $y$-directions are given by replacing 
$m^{2}_{y}$ with $m^{2}_{y}\mp n\partial_\tau B$.
In this case, we obtain the Mathieu equations with 
\begin{equation}
C_{2}=-{1\over 2}(1+3\cos^2\gamma)\tilde{Q}_{5}\pm 
2n\ell\cos\gamma\tilde{Q}_{5}\qquad  
\mbox{for $\varphi_{n}^{(\pm)}$ ($y$-directions)}.
\end{equation}
The NS5-brane case is given by setting $\cos\gamma=1$
(and replacing $\tilde{Q}_{5}$ with $Q_{5}$)
in the above expressions.

\subsection{Exact solutions}
For the equation of motion in the $y$-directions,
we can find the exact solutions. 
First, we consider the D5-brane case.
The equation of motion is given by (\ref{DEoddp})-(\ref{eq:defqx})
with $m^{2}_{y}$ in (\ref{eq:myD5}) and $Q(r)$ in (\ref{eq:QrD5}).
With the substitution
$\varphi_{n}(x)=f(x)\phi_{n}(x)$,
we obtain
\begin{equation}
\left( {d^2\over dx^2}
+\tilde{p}(x){d\over dx}+\tilde{q}(x)
\right)\phi_{n}(x)=0,
\label{eq:schroe4}
\end{equation}
where
\begin{eqnarray}
\tilde{p}(x)&=&p(x)+{2f'(x)\over f(x)}, \nonumber \\
\tilde{q}(x)&=& q(x)+{f'(x)\over f(x)}p(x)+{f''(x)\over f(x)}.
\end{eqnarray}
Since $p(x)$ and $q(x)$ have the singular points at 
$x=0$, $-Q_{5}$ and $-Q_{5}+\ell^2$, we will use 
$f(x)=x^{\alpha}(x+Q_{5})^{\beta}(x+Q_{5}-\ell^2)^{\gamma}$.
Then $\tilde{q}(x)$ takes the form 
$q_{1}(x)/(x^2(x+Q_{5})^2(x+Q_{5}-\ell^2)^2)$, 
where $q_{1}(x)$ is the fifth
order polynomial in $x$.
We choose the exponents $(\alpha,\beta,\gamma)$ 
such that $q_{1}(x)=0$
for $x=0$, $-Q_{5}$ and $-Q_{5}+\ell^2$. 
These conditions determine the exponents as follows:
\begin{eqnarray}
\alpha&=&\alpha_{\pm}\equiv {1\over4}\pm {1\over2}
\left(\ell^2\over \ell^2-Q_{5}\right)^{1\over2},\nonumber\\
\beta&=&{1\over4}, {5\over4},\nonumber\\
\gamma&=&0,{1\over2}.
\end{eqnarray}
The choice $(\alpha,\beta,\gamma)=(\alpha_{+},{1\over4},0)$ takes
the functions
$\tilde{p}(x)$ and $\tilde{q}(x)$
 into simpler forms:
\begin{eqnarray}
 \tilde{p}(x)&=&{{1\over2}+2\alpha_{+}\over x}
+{{1\over2}\over x+Q_{5}-\ell^2},\nonumber\\
\tilde{q}(x)&=&
{1\over4}{n^2 Q_{5}(\ell^2-Q_{5})+\ell^2+\sqrt{\ell^2(\ell^2-Q_{5})}
+n^2(\ell^2-Q_{5})x\over x(x+Q_{5}-\ell^2)(\ell^2-Q_{5})}.
\label{eq:coeff1}
\end{eqnarray}
We notice that the present form of the differential equation 
(\ref{eq:schroe4}) with (\ref{eq:coeff1}) is nothing but the associated
Mathieu equation \cite{In}. 
In fact, by the substitution $x=(Q_{5}-\ell^2)(z-1)$, we get
\begin{equation}
 {d^2\phi_{n}\over dz^2}+
\left\{
{{1\over2}\over z}+{1-r\over z-1}
\right\}{d\phi_{n}\over dz}
-{a+k^2 z\over 4z(z-1)}\phi_{n}=0,
\label{eq:assmat}
\end{equation}
where
\begin{eqnarray}
r&=& {1\over2}-2\alpha_{+}
=- \left({\ell^2\over \ell^2-Q_{5}}\right)^{1\over2},
\nonumber\\
k^2&=& n^2(\ell^2-Q_{5}),\\
a&=&{-n^2\ell^2(\ell^2-Q_{5})-\ell^2-\sqrt{\ell^2(\ell^2-Q_{5})}\over
 \ell^2-Q_{5}}.
\end{eqnarray}
It is known that the associated Mathieu equation (\ref{eq:assmat}) 
is equivalent to  the
differential equation for the spheroidal wave function 
\cite{Er} by the 
transformation $w=z^{1\over2}$ and $\phi_{n}= (1-w^2)^{r\over2} f_{n}$:
\begin{equation}
(1-w^2) {d^2f_{n}\over dw^2}-2w{df_{n}\over dw}+\left(\lambda+4\theta (1-w^2)
-{\mu^2\over 1-w^2}\right)f_{n}=0,
\label{eq:sphe}
\end{equation}
where
\begin{eqnarray}
\lambda&=& a+r(r-1)+k^2=-n^2Q_{5},\nonumber\\
 \theta&=& -{k^2\over 4}=-{n^2(\ell^2-Q_{5})\over 4},\nonumber\\
 \mu&=& r=-\sqrt{\ell^2\over \ell^2-Q_{5}}.
\end{eqnarray}
The regular singular points $x=\ell^2-Q_{5}$ and $x=0$ correspond to $w=0$
and $w=\pm 1$, respectively.
The irregular singular point at $x=\infty$ corresponds to $w=\infty$.

The spheroidal wave function with order $\mu$ has been investigated in 
refs.~\cite{Er,MeSc}. 
With the replacement $\zeta=2\theta^{1/2}w$, (\ref{eq:sphe}) becomes
\begin{equation}
 (\zeta^2-4\theta){d^2 f_{n}\over d\zeta^2}+2\zeta{df_{n}\over d\zeta}
+\left(\zeta^2-\lambda-4\theta-{4\theta\mu^2\over \zeta^2-4\theta}\right)
f_{n}=0.
\label{eq:sphe1}
\end{equation}
In the $\theta=0$ limit, we obtain the Bessel equation which have two 
independent solutions:
\begin{equation}
 \psi^{(1)}_{\nu}(\zeta)=
\left({\pi\over2\zeta}\right)^{1/2}J_{\nu+{1\over2}}(\zeta),\quad
 \psi^{(2)}_{\nu}(\zeta)=
\left({\pi\over2\zeta}\right)^{1/2}Y_{\nu+{1\over2}}(\zeta),\quad
\end{equation}
where $\lambda=\nu(\nu+1)$. $J_{\nu+1/2}(\zeta)$ is the Bessel function
and $Y_{\nu+1/2}(\zeta)$ is the Neumann function.
One may also introduce the Hankel functions by
\begin{equation}
 \psi^{(3)}_{\nu}=\psi^{(1)}_{\nu}+i\psi^{(2)}_{\nu}=
\left({\pi\over2\zeta}\right)^{1/2}H^{(1)}_{\nu+{1\over2}}(\zeta),\quad
 \psi^{(4)}_{\nu}=\psi^{(1)}_{\nu}-i\psi^{(2)}_{\nu}=
\left({\pi\over2\zeta}\right)^{1/2}H^{(2)}_{\nu+{1\over2}}(\zeta).
\end{equation}
The spheroidal wave functions are given by the
expansions
of the form
\begin{equation}
 S^{\mu(j)}_{\nu}(w,\theta)=(1-w^{-2})^{-\mu/2}s^{\mu}_{\nu}(\theta)
\sum_{r=-\infty}^{\infty}a^{\mu}_{\nu,r}(\theta)\psi^{(j)}_{\nu+2r}
(2\theta^{1/2}w),\quad
(j=1,2,3,4)
\end{equation}
where $s^{\mu}_{\nu}(\theta)$ is a normalization constant and 
$a^{\mu}_{\nu,r}(\theta)$ obey a recursion relation:
\begin{eqnarray}
&& {(\nu+2r-\mu)(\nu+2r-\mu-1)\over
(\nu+2r-3/2)(\nu+2r-1/2)}\theta a^{\mu}_{\nu,r-1}(\theta)
+
 {(\nu+2r+\mu+2)(\nu+2r+\mu+1)\over
(\nu+2r+3/2)(\nu+2r+5/2)}\theta a^{\mu}_{\nu,r+1}(\theta)\nonumber\\
&&+\left[
\lambda-(\nu+2r)(\nu+2r+1)+{(\nu+2r)(\nu+2r+1)+\mu^2-1\over
(\nu+2r-1/2)(\nu+2r+3/2)} 2\theta
\right] a^{\mu}_{\nu,r}(\theta)=0
\end{eqnarray}

Asymptotic behavior of $S^{(j)}$ as $w\rightarrow\infty$ has been 
determined by Meixner \cite{Er,MeSc}. 
If we choose the normalization factor $s^{\mu}_{\nu}(\theta)$ as
\begin{equation}
 s^{\mu}_{\nu}(\theta)={1\over \sum_{r=-\infty}^{\infty}(-1)^{r}
a^{\mu}_{\nu,r}(\theta)},
\end{equation}
the asymptotic forms may be written as
\begin{equation}
 S^{\mu(j)}_{\nu}(w,\theta)\sim \psi^{(j)}_{\nu}(2\theta^{1/2}w),
\quad w\rightarrow\infty, \quad (j=1,2,3,4)
\end{equation}
where $|{\rm arg}(\theta^{1/2}w)|<\pi$.
In particular, using the asymptotic expansion of the Hankel functions
$H^{(i)}_{\nu+1/2}$ ($i=1,2$), the asymptotic expansions of 
$S^{\mu (3)}_{\nu}(w,\theta)$
and
$S^{\mu (4)}_{\nu}(w,\theta)$ for large $|w|$ are given by
\begin{eqnarray}
 S^{\mu (3)}_{\nu}(w,\theta)&=&{1\over 2\theta^{1/2}w}
e^{i (2\theta^{1/2}w-\nu\pi/2-\pi/2)}\left[1+O(|w|^{-1})\right],\quad
w\rightarrow\infty, -\pi<{\rm arg}(\theta^{1/2}w)<2\pi\nonumber\\
 S^{\mu (4)}_{\nu}(w,\theta)&=&{1\over 2\theta^{1/2}w}
e^{-i (2\theta^{1/2}w-\nu\pi/2-\pi/2)}\left[1+O(|w|^{-1})\right],\quad
w\rightarrow\infty, -2\pi<{\rm arg}(\theta^{1/2}w)<\pi.\nonumber\\
\end{eqnarray}
Let us go back to the equation (\ref{eq:schroe}) in the
$y$-direction. 
We are interested in the two solutions $\varphi_{n}(\tau)$ and
$\tilde{\varphi}_{-n}(\tau)$
whose asymptotic behaviors for $\tau\rightarrow\infty$ 
take the form
\begin{eqnarray}
 \varphi_{n}(\tau)\sim e^{in\tau},\nonumber\\
 \tilde{\varphi}_{-n}(\tau)\sim e^{-in\tau}.
\label{eq:asympt}
\end{eqnarray}
Since $\theta=\sqrt{n^2(Q_{5}-\ell^2)}/2$ and 
$w=\left(1+{x\over Q_{5}-\ell^2}\right)^{1\over2}$, 
we find
\begin{equation}
2\theta^{1/2}w=\sqrt{n^2(x-\ell^2+Q_5)}\sim |n|\tau,
 \quad (\tau\rightarrow\infty)
\end{equation}
where we have used $x=r^2\sim \tau^2$ as
$r\rightarrow\infty$.
Therefore using the spheroidal wave functions
$S^{\mu (3)}_{\nu}(w,\theta)$
and
$S^{\mu (4)}_{\nu}(w,\theta)$, we express the solutions of 
(\ref{eq:schroe}) as
\begin{eqnarray}
\hspace*{-1cm}
 \varphi_{n}(\tau)=\tilde{\varphi}_{n}(\tau)&\!=\!&
\left\{
\begin{array}{l}
A_{n}x^{1/4}(x+Q_{5})^{1/4}
S^{\mu (3)}_{\nu}(\sqrt{n^2(x-\ell^2+Q_5)},\,\theta)
\qquad (n>0), \\
A_{n}^{\ast}x^{1/4}(x+Q_{5})^{1/4}
S^{\mu (4)}_{\nu}(\sqrt{n^2(x-\ell^2+Q_5)},\,\theta)\qquad (n<0)
\end{array}
\right. 
\label{eq:phis1}
\end{eqnarray}
where 
\begin{equation}
 A_{n}=|n|e^{i(\nu\pi/2+\pi/2)}.
\label{eq:AB}
\end{equation}

Similar analysis can be done in the case of the 
$(p,q)$ fivebranes as well as the NS5-branes.
The string equation of motion in the $y$-directions are given by
\begin{equation}
\Big( \partial^2_\tau +n^2 -m_{y}(\tau)^2
\pm n \partial_\tau B(\tau) \Big) \varphi^{(\pm)}_{n}=0
\label{eq:pqsch}
\end{equation}
where $m_{y}^2(\tau)$ is in (\ref{pq5my}) and $\partial_{\tau} B(\tau)$ is in
(\ref{eq:tBpq5}).
As in the case of the D5 branes, we change variable $\tau$ into 
$x=r^2$
and $\varphi^{(\pm)}_{n}(\tau)=f(x)\phi^{(\pm)}_{n}(x)$, where
$f(x)=x^{\alpha}(x+\tilde{Q}_{5})^{\beta}(x+\tilde{Q}_{5}-\ell^2)^{\delta}
(x+\tilde{Q}_{5}\cos^2\gamma)^{\eta}$.
The choice of the exponents $(\alpha,\beta,\delta,\eta)=
(\alpha_{+}^{(\pm)},{1\over4},0,{1\over4})$ 
simplify the differential equation (\ref{eq:pqsch}),where
\begin{equation}
 \alpha_{+}^{(\pm)}={1\over2}{\ell \mp n \tilde{Q}_{5}\cos\gamma\over \sqrt{
\ell^2-\tilde{Q}_{5}}}.
\end{equation}
$\phi^{(\pm)}_{n}(x)$ satisfy the equation (\ref{eq:schroe4}) with
\begin{eqnarray} 
 \tilde{p}(x)&=& {1+2\alpha_{+}^{(\pm)}\over x}
+{{1\over2}\over x+\tilde{Q}_{5}-\ell^2 }\\
\tilde{q}(x)&=&
-{1\over 4x (x+\tilde{Q}_{5}-\ell^2) (\ell^2-\tilde{Q}_{5})}\Big(
n^2\tilde{Q}_{5}(\tilde{Q}_{5}-\ell^2-\cos^2\gamma\ell^2)-\ell^2
\nonumber\\
&&\quad
\pm 2n\tilde{Q}_5\ell\cos\gamma-
\sqrt{\ell^2-\tilde{Q}_{5}}(\ell\mp n\tilde{Q}_5\cos\gamma)
-n^2(\ell^2-\tilde{Q}_{5})x
\Big).
\label{eq:coeff2}
\end{eqnarray}
This differential equation also becomes the associated Mathieu equation
(\ref{eq:assmat})
by the change of the variables $x=(\tilde{Q}_{5}-\ell^2)(z-1)$, 
where
\begin{eqnarray}
 r&=& -2\alpha^{(\pm)}_{+}= 
{\ell \mp n \tilde{Q}_{5}\cos\gamma\over \sqrt{
\ell^2-\tilde{Q}_{5}}},\nonumber\\
k^2&=& n^2(\ell^2-\tilde{Q}_{5}),\nonumber\\
a&=& {1\over \ell^2-\tilde{Q}_{5}}
\Bigl(n^2\ell^2(\tilde{Q}_{5}-\ell^2-\tilde{Q}_{5}\cos^2\gamma)-\ell^2
\pm 2n\tilde{Q}_{5}\ell \cos\gamma
\nonumber\\
&& -\sqrt{\ell^2-\tilde{Q}_{5}}(\ell\mp n\tilde{Q}_{5}\cos\gamma)
\Bigr).
\end{eqnarray}
By the transformation
$w=z^{1/2}$ and $\phi^{(\pm)}_{n}=(1-w^2)^{r/2}f^{(\pm)}_{n}$,
we obtain the differential equation for the spheroidal wave functions
(\ref{eq:sphe}) with parameters
\begin{eqnarray}
 \lambda&=& -n^2\tilde{Q}_{5}(1+\cos^2\gamma),\nonumber\\
 \theta&=&-{n^2(\ell^2-\tilde{Q}_{5})\over 4},\nonumber\\
\mu&=&  
{\ell \mp n \tilde{Q}_{5}\cos\gamma\over \sqrt{
\ell^2-\tilde{Q}_{5}}}.
\end{eqnarray}
By using the spheroidal wave function $S^{\mu (j)}_{\nu}(w,\theta)$
with $\lambda=\nu(\nu+1)$,
we may solve 
the wave functions $\varphi_{n}^{(\pm)}(\tau)$ and 
$\tilde{\varphi}_{-n}^{(\pm)}(\tau)$
with asymptotic behavior (\ref{eq:asympt}).
These are expressed as
\begin{eqnarray}
&& \varphi^{(\pm)}_{n}(\tau)=\tilde{\varphi}^{(\pm)}_{n}(\tau) 
\nonumber \\
&&=
\left\{
\begin{array}{l}
A_{n}(x+Q_{5})^{1/4}
(x+\tilde{Q}_{5}\cos^2\gamma)^{1/4}
S^{\mu (3)}_{\nu}(\sqrt{n^2(x-\ell^2+\tilde{Q}_5)},\,\theta)\qquad (n>0),\\
A_{n}^{\ast}(x+Q_{5})^{1/4}(x+\tilde{Q}_{5}\cos^2\gamma)^{1/4}
S^{\mu (4)}_{\nu}(\sqrt{n^2(x-\ell^2+\tilde{Q}_5)},\,\theta)\qquad (n<0)
\end{array}
\right.\nonumber\\
\end{eqnarray}
where the constant
 $A_{n}$ is the same as (\ref{eq:AB}).

The solution at $w=0$ corresponds to the theory at the critical
radius. The expansion of the spheroidal wave functions 
in the region $|w\pm 1|<2$ are written in terms of the
Legendre polynomials \cite{Er,MeSc}, but in this paper,
we do not discuss this expansion.

\section{Conclusions and discussion}
In this paper, we have studied string theories on the Penrose 
limit of various brane solutions of type II supergravity.
Firstly, we obtained the plane wave geometry for the D$p$-brane,
NS and $(p,q)$ fivebrane, fundamental and $(p,q)$ string
solutions, not restricting ourselves to the 
near-horizon limit. 
We have thoroughly investigated the equations of motion of 
the light-cone bosonic string on these backgrounds, which have
time-dependent masses. 

We have found several types of the solutions. 
The simplest class of the solutions are given when
the masses become constant. These arise 
when we consider the Penrose limit along a geodesic 
which stays at a fixed radius. We have analyzed the 
condition for the existence of this kind of geodesics,
and gave the formula for the mass terms.
In the near-horizon limit, we have various examples
in which the equations of motion can be solved 
by the Bessel functions. We presented a preliminary
analysis for the canonical quantization of the string
for the latter case.

We found non-trivial solutions for the case
of fivebrane backgrounds. Firstly, 
the equations of motion can be solved by the 
Mathieu functions in the region near the 
critical radius (the point on the geodesic 
which is nearest to the origin). 
In the asymptotic region at infinity,
we found solutions also in terms of the Mathieu
functions. Furthermore, for a particular component,
exact solutions exist which is written by using
the spheroidal wave functions. The above solutions
exist for the fivebranes with general $(p,q)$.

For physical applications, it is important to extend
our analysis to the fermionic sector of the superstring.
Green-Schwarz action for the superstring can be written in 
arbitrary background in principle \cite{Gr}.
It is a non-trivial question whether the equation of
motion for fermionic sectors derived from that action
can be solved as we have done for the bosonic sectors.
It is also interesting to attempt a study in the NSR
formalism. In some of our plane wave backgrounds (for NS branes,
or for the case of radial null geodesics), RR backgrounds
are not present. For these cases, the latter formalism
may be more useful.

Most interesting problem for the future study is the 
quantization of the string on the plane wave backgrounds
for the brane solutions. As discussed briefly in the appendix,
performing the mode expansion using the solutions of 
the equations of motion as the  basis, and following 
the procedure of the canonical quantization,
we can obtain the oscillators and construct the Fock space. 
For time-dependent plane wave backgrounds, such states 
do not diagonalize the light-cone Hamiltonian in general. 
In other words, an asymptotic state will evolve into another due
to the influence of the brane background.
For the fivebrane backgrounds we can cover
the whole region of the geodesics either by matching the
solutions obtained in the two limits, or by using 
the exact solution. It should enable us to study the 
scattering or absorption of string states by the fivebranes.
We hope to report progress in this direction
in a future publication.

It would also be interesting to  apply our results to the 
holographic correspondence with gauge theories. 
As mentioned in the introduction,
six-dimensional gauge theory is investigated from
the perspective of the Penrose limit of the NS5-brane 
background \cite{HuRaVe,OzSa}. The results of this paper
such as the classification of  the geodesics and the exact 
solutions for the fivebrane backgrounds
may give further information on the gauge theory.

Holographic dualities for general (dilatonic) 
D$p$-branes have been proposed by  
Itzhaki et.al. \cite{IzMaSoYa},
but they are less understood than the $AdS_5/CFT_4$ 
correspondence. There are few
quantitative studies on those dualities.
(See refs.~\cite{SeYo, Se} for such
an attempt for the case of the D0-branes.)
It would be nice if it is possible to 
calculate the gauge theory correlators from string 
theory on the Penrose limit of
the brane backgrounds.
Our solution of the string equations of motion
in the near-horizon limit of D$p$-branes (with
$p\le4$) may be helpful in this regard. 
We have found the solutions
taking an additional limit (\ref{nhlimit2}), but
further analysis without that restriction would
be important. 

Holographic renormalization group for nonconformal field 
theories is also an interesting subject for future studies.
Analysis of the holographic RG for the pp-wave backgrounds 
has been done in refs.~\cite{CoHaKeWa, GiPaSo, BrJoLoMy},
where the radial coordinate plays the role of the scale of 
the dual gauge theories. It is an interesting question
how to interpret the string theories on the critical
radius in this context.

\vspace{1cm}

\noindent{{\bf Acknowledgement}}

The work of H.F. is supported 
by JSPS research fellowship for young scientists.

\section*{Appendix: Mode expansion using the Hankel functions}
\renewcommand{\theequation}{A.\arabic{equation}}
\setcounter{equation}{0}
In this appendix, we give a preliminary analysis for the canonical
quantization of the string on a particular plane wave background.
We expand $X^{i}$ using the solutions of the equations
of motion as the basis and impose the canonical 
commutation relation.
We demonstrate that we can construct the oscillators,
and present the form of the light-cone Hamiltonian.

Consider the string equation of motion of the form
\begin{equation}
\left({d^2\over d\tau^2}+n^2
- {C_{1}\over \tau^{2}} \right)\varphi_{n}=0.
\label{eq:appeomBessel1}
\end{equation}
This equation is obtained when we study the near-horizon 
limit of various branes, as we have seen in section 4.3. 
The constant $C_{1}$ depends on the model. 
We shall solve (\ref{eq:appeomBessel1})
with the following boundary condition for 
$\tau\rightarrow\infty$:
\begin{equation}
\varphi_{n}(\tau)\rightarrow e^{in\tau},\quad
\tilde{\varphi}_{-n}(\tau)\rightarrow e^{-in\tau}.
\label{eq:assbc}
\end{equation}
Note that (\ref{eq:appeomBessel1}) approaches the equation
of motion in the flat background asymptotically
(as $\tau\rightarrow\infty$).
The above boundary condition is the one which 
match the mode expansion in the flat background\footnote{Since 
the present theory is obtained
by taking the near-horizon limit and is not valid for 
$r\rightarrow\infty$ (or $\tau\rightarrow\infty$),
whether we should choose this boundary condition is not clear.}.

The solutions for the non-zero modes are given by 
\begin{eqnarray}
&&\varphi_{n}(\tau)=\tilde{\varphi}_{n}(\tau)=\left\{
\begin{array}{l}
\sqrt{{\pi\over 2}}e^{{i\over 2}(\nu +{1\over 2})\pi}
(n\tau)^{{1\over 2}}H^{(1)}_{\nu}(n\tau)\qquad\qquad (n>0)\\
\sqrt{{\pi\over 2}}e^{-{i\over 2}(\nu +{1\over 2})\pi}
(-n\tau)^{{1\over 2}}H^{(2)}_{\nu}(-n\tau)\qquad (n<0)\\
\end{array}\right.
\end{eqnarray}
where $\nu^2\equiv C_{1}+1/4$ and 
$H^{(1)}_{\nu}$ and $H^{(2)}_{\nu}$ are the Hankel functions
which are related to the Bessel functions of the first kind
$J_{\nu}$ and the second kind $Y_{\nu}$ by
\[
H^{(1)}_{\nu}(x)=J_{\nu}(x)+iY_{\nu}(x),\quad
H^{(2)}_{\nu}(x)=J_{\nu}(x)-iY_{\nu}(x).
\]
The two independent solutions for the zero-modes
are 
\begin{equation}
\varphi_0=\tau^{\nu +{1\over 2}},\qquad
\tilde{\varphi_0}=\tau^{-\nu +{1\over 2}}.
\end{equation}

We perform the quantization of the string using 
the solutions found above as the basis functions.
Remember that $X^{i}$ is given as
\[
X^{i}(\tau,\sigma)=\sum_{n}\left(\alpha_{n}^{i}
\varphi_{n}(\tau)+ \tilde{\alpha}_{-n}^{i}
\tilde{\varphi}_{-n}(\tau) \right)
e^{in\sigma}.
\]
The reality condition for $X^i$ leads to
\begin{equation}
\tilde{\alpha}^{i}_{n}{}^{\dagger}=\tilde{\alpha}^{i}_{-n},\quad
\alpha^{i}_{n}{}^{\dagger}=\alpha^{i}_{-n},
\end{equation}
where we have used
$H^{(1)}{}^*(x) =H^{(2)}(x)$ for a real $x$.

Now we impose the canonical commutation relation for $X^i$ 
\begin{equation}
[X^{i}(\tau,\sigma),\partial_\tau X^{j}(\tau,\sigma')]
=-i 2\pi\alpha' \delta^{ij} \delta(\sigma-\sigma').
\label{CRX}
\end{equation}
This implies the following
commutation relations for the oscillators
\begin{eqnarray}
[\tilde{\alpha}^{i}_{n},\tilde{\alpha}^{j}_{n'}]
&=&{1\over 2n}\delta^{ij}\delta_{n,-n'}\qquad (n\neq 0),\nonumber\\
{}[\alpha^{i}_{n},\alpha{}^{j}_{n'}]
&=&{1\over 2n}\delta^{ij}\delta_{n,-n'}\qquad (n\neq 0),\nonumber\\
{}[\alpha^{i}_{0},\tilde{\alpha}^{j}_{0}]&=&i\delta^{ij},
\label{CRalpha}
\end{eqnarray}
where commutators among other components are zero.
The consistency between (\ref{CRX}) and  (\ref{CRalpha}) 
is guaranteed by the following equation (for the non-zero
mode part) 
\begin{eqnarray}
\varphi_{n}\partial_{\tau}\varphi_{-n}
-\tilde{\varphi}_{-n}\partial_{\tau}\tilde{\varphi}_{n}
&=&\varphi_{n}\partial_{\tau}\varphi_{-n}
-\varphi_{-n}\partial_{\tau}\varphi_{n}\\
&=&{\pi n\tau\over 2}\left( 
H^{(1)}_{\nu}(n\tau)\partial_{\tau}H^{(2)}_{\nu}(n\tau)
-H^{(2)}_{\nu}(n\tau)\partial_{\tau}H^{(1)}_{\nu}(n\tau)\right)
=-{2ni}\nonumber
\label{eq:lommel}
\end{eqnarray}
where the last equality is due to the Lommel's formula 
for the Hankel functions. 

Note that 
\begin{equation}
\varphi_{n}\partial_{\tau}
\varphi_{-n}-\varphi_{-n}\partial_{\tau}\varphi_{n}=2ni
\label{eq:wronskian}
\end{equation}
is the Wronskian of the two independent solutions
$\varphi_{n}$ and $\varphi_{-n}$ of (\ref{eq:appeomBessel1}). 
We can easily prove that the Wronskian for our 
differential equation is $\tau$-independent,
by taking its derivative and using (\ref{eq:appeomBessel1}).
Thus, (\ref{eq:wronskian}) is given by the value evaluated
at $\tau\rightarrow \infty$, which is the r.h.s. of that
equation.
Moreover, the Wronskian for a differential equation of 
the form (\ref{DE1}) is independent of $\tau$ 
for arbitrary $m^{2}_{i}(\tau)$. 
Thus, if we consider an equation with $m^{2}_{i}(\tau)\rightarrow 
0$ as $\tau\rightarrow\infty$, and if we take the 
solutions with the boundary condition (\ref{eq:assbc}), 
the value of the Wronskian agrees with (\ref{eq:wronskian}).
This fact suggests that we can get
the commutation relations for the oscillators 
(\ref{CRalpha}) from the canonical commutation
relation (\ref{CRX}), for those generic plane wave backgrounds.

The light-cone Hamiltonian for the model of 
(\ref{eq:appeomBessel1}) is given by
\begin{equation}
H={1\over 2\pi\alpha'P^{+}}\int d\sigma
\left( {1\over 2}(\partial_{\tau}X^{i})^2 
+{1\over 2}(\partial_{\sigma}X^{i})^2
+{C_{1}\over 2\tau^2}(X^{i})^2\right).
\end{equation}
Substituting the mode expansion, we obtain
\begin{eqnarray}
H&=&{1\over 2\alpha'P^{+}}
\sum_{n}\left( (\partial_{\tau}\varphi_{n}
\partial_{\tau}\varphi_{-n} +(n^2+{C_{1}\over \tau^2})
\varphi_{n}\varphi_{-n}) \alpha^{i}_{n}\alpha^{i}_{-n}\right.
\nonumber\\
&&\qquad\qquad +(\partial_{\tau}\tilde{\varphi}_{n}
\partial_{\tau}\tilde{\varphi}_{-n} +(n^2+{C_{1}\over \tau^2})
\tilde{\varphi}_{n}\tilde{\varphi}_{-n}) 
\tilde{\alpha}^{i}_{n}\tilde{\alpha}^{i}_{-n}\nonumber\\
&&\qquad\qquad \left. +2(\partial_{\tau}\varphi_{n}
\partial_{\tau}\tilde{\varphi}_{n} +(n^2+{C_{1}\over \tau^2})
\varphi_{n}\tilde{\varphi}_{n}) 
\alpha^{i}_{n}\tilde{\alpha}^{i}_{n} \right).
\label{eq:LChamil}
\end{eqnarray}
The Hamiltonian is time dependent, and especially,
it does not commute with the number operator.
The first two terms conserve the particle number,
but the last term does not.
Note that the case of the flat background is given by
substituting $C_{1}=0$ and 
$\varphi_{n}=\tilde{\varphi}_{n}=e^{in\tau}$ into 
(\ref{eq:LChamil}). In this case, the last term 
of (\ref{eq:LChamil}) vanish,
but for the time-dependent plane wave backgrounds,
that term should be present in general.

\end{document}